\documentclass[journal,twoside,web]{ieeecolor}
\usepackage{generic}
\usepackage{cite}
\usepackage{amsmath,amssymb,amsfonts}
\usepackage{algorithmic}
\usepackage{graphicx}
\usepackage{textcomp}
\usepackage{float}
\usepackage{subfigure}
\usepackage[justification=centering]{caption}
\usepackage{bm}
\newtheorem{theorem}{Theorem}[section]

\newtheorem{definition}{Definition}[section]
\newtheorem{assumption}{Assumption}[section]

\newtheorem{lemmax}{Lemma}[section]
\newtheorem{remark}{Remark}[section]
\def\BibTeX{{\rm B\kern-.05em{\sc i\kern-.025em b}\kern-.08em
    T\kern-.1667em\lower.7ex\hbox{E}\kern-.125emX}}
\begin{document}
\title{\bf Differentially Private Bipartite Consensus over Signed Networks with Time-Varying Noises
\thanks{The work was supported by National Key R\&D Program of China under Grant 2018YFA0703800, National Natural Science Foundation of China under Grant 61877057, and China Post-Doctoral Science Foundation under Grant 2018M641506. The material in this paper was not presented at any conference. Corresponding author: Ji-Feng Zhang.}}
\author{Jimin Wang, \IEEEmembership{Member,~IEEE,} Jieming Ke and Ji-Feng Zhang, \IEEEmembership{Fellow,~IEEE}
\thanks{Jimin Wang is with the School of Automation and Electrical Engineering, University of Science and Technology Beijing, Beijing, 100083, China (e-mails: jimin.wang@amss.ac.cn)}
\thanks{Jieming Ke and Ji-Feng Zhang are with the Key Laboratory of Systems and
Control, Institute of Systems Science, Academy of Mathematics and
Systems Science, Chinese Academy of Sciences, Beijing 100190, and also with the School of Mathematical Sciences, University of Chinese Academy of Sciences, Beijing 100049, P. R. China. (e-mails: jif@iss.ac.cn)}}

\maketitle
\begin{abstract}
This paper investigates the differentially private bipartite consensus algorithm over signed networks. The proposed algorithm protects each agent's sensitive information by adding  noise with time-varying variances to the cooperative-competitive interactive information. In order to achieve privacy protection, the variance of the added noise is allowed to be increased, and substantially different from the existing
works. In addition, the variance of the added noise can be either decaying or constant. By using time-varying step-sizes based on the stochastic approximation method, we show that the algorithm converges in mean-square and almost-surely even with an increasing privacy noise. We further develop a method to design the step-size and the noise parameter, affording the algorithm to achieve asymptotically unbiased bipartite consensus with the desired accuracy and the predefined differential privacy level. Moreover, we give the mean-square and almost-sure convergence rate of the algorithm, and the privacy level with different forms of the privacy noises. We also reveal the algorithm's trade-off between the convergence rate and the privacy level. Finally, a numerical example verifies the theoretical results and demonstrates the algorithm's superiority against existing methods.
\end{abstract}

\IEEEpeerreviewmaketitle

\section{Introduction}
\IEEEPARstart{D}{istributed} consensus control of multi-agent systems (MASs) is significant due to its numerous applications, such as energy internet \cite{Zhang20121761AveConsensusinES,Chen2021a}, cooperative guidance systems \cite{Kang2018GuidanceSystem}, and social networks \cite{Wang2021}. Generally, it refers to designing a network protocol such that all agents asymptotically reach an agreement with time. To date, many works have been developed on the consensus control of MASs, including average consensus \cite{Olfati2004Consensus,Li2009,Li2010CommunicationNoise,Huang2010,Huang2012,Huang2015}, min-max consensus \cite{Mulla2018}, group consensus \cite{Chen2019, Li2022}, and bipartite consensus \cite{Altafini2013,Liu2017,Hu2019,Chen2021b}. Among others, cooperative and competitive interactions exist simultaneously in many complex MASs \cite{Shi2016,Shi2019}. For modeling such networks, signed graph theory and bipartite consensus problems were formulated in \cite{Altafini2013}, where the agents achieved an agreement with identical values but opposite signs. Currently, some substantial progresses have been made for bipartite consensus control of MASs.

With the increasing need for privacy and security, preserving the privacy of an individual dataset is required in many applications. For example, in a social network \cite{Amelkin2017OpinionSystem}, people interact with their neighbors, and evaluate their opinions in comparison with the others' opinions. However, exchanging opinions probably reveals individual privacy. In this case, privacy-preserving in social networks has become a hot research topic. In a cooperative guidance system \cite{Kang2018GuidanceSystem},  information interactions may expose the missile's and the launch stations' location. Hence, a naturally arising problem is how to achieve a bipartite consensus while protecting each agent's sensitive information from being inferred by potential attackers.

To address the requirement for privacy protection in distributed control, some methods have been proposed recently to counteract such potential privacy breaches, such as homomorphic encryption \cite{Ruan2019encryption,Lu2018}, adding noise \cite{Mo2017CNPP,He2019CNPPEX,He2019CNPP}, time-varying transformation \cite{Altafini2020,Wang2022a}, and state decomposition \cite{Wang2019}. Homomorphic encryption allows direct calculation of encrypted data without revealing any information about the original text. However, such approaches incur a heavy communication and computation overhead. Accurate consensus can be achieved by adding correlated noises to interaction information  while protecting the initial states from semi-honest agents \cite{Mo2017CNPP,He2019CNPPEX,He2019CNPP}. However, if the potential passive attackers obtain the information received and delivered by agent $i$, then this agent's initial state can be estimated through an iterative observer under such correlated noise mechanism. Generally, current methods considering privacy preservation in average consensus assume that the honest-but-curious adversary cannot access the entire neighborhood set of an agent \cite{Ruan2019encryption,Mo2017CNPP,Wang2019,Altafini2020}.

Differential privacy techniques have been widely considered when publishing data from many technology companies, such as Google and Apple. Based on the original definition given by \cite{Dwork2006DP}, $\epsilon$-differential privacy has been extended to a multi-agent scenario, including protecting the initial agent states in a consensus problem \cite{Huang2012DP}, protecting the objection function in distributed optimization \cite{Han2017,Wang2022b} and games \cite{WangJM2022}, and protecting the global state trajectories in Kalman filtering \cite{Ny2014DP,Ny2018DP}. From a system control perspective, a tutorial and comprehensive framework of privacy security on control systems is provided in \cite{Zhang2021}. By adding uncorrelated noises on information, a differentially private consensus algorithm is designed for discrete-time multi-agent systems, where agents achieved unbiased convergence to the average almost-surely \cite{Nozari2017Differentially,Fiore2019DP}. An $\epsilon$-differentially private consensus algorithm is designed in \cite{Liu2020} for continuous-time heterogeneous multi-agent systems, while an event-triggered scheme is proposed in \cite{Gao2019DP} to reduce the control updates and the $\epsilon$-differentially private of the algorithm is ensured. Overall, current literature has two common grounds: 1) all algorithms are designed for average consensus, and 2) in order to guarantee the convergence and satisfy the differential privacy level, the privacy noise is required to be decaying exponentially to zero (or constant) with time. In fact, the interaction between practical systems involves cooperation and competitiveness simultaneously. Although the differential privacy bipartite consensus over signed graphs is considered, the privacy noise with exponential decay to zero is required in \cite{Zuo2022}. Note that decaying noises to zero potentially expose the trajectory of the state. Then, the following questions arise. Is it possible to give a more general noise form for privacy-preserving distributed consensus algorithm? If possible, how does the added privacy noise affect the algorithm's convergence rate and privacy level? These questions motivate us to investigate the privacy-preserving bipartite consensus algorithm and relax the limitation of the existing privacy noise forms.

This paper designs a distributed differentially private bipartite consensus algorithm over signed networks. Specifically, each agent adds a Laplace noise on the local state, and then transmits it to its neighbors. The added noises are i.i.d. with a time-varying variance (which may increase with time). If the algorithm's step-size $\alpha(k)$ satisfies the stochastic approximation condition, then the algorithm can achieve asymptotically unbiased mean-square and almost-sure bipartite consensus. In summary, the contributions of this paper are threefold:
\begin{itemize}
\item A differentially private bipartite consensus algorithm is developed, compared with existing literature that focuses on bipartite consensus \cite{Altafini2013,Liu2017,Hu2019,Chen2021b}. In order to achieve privacy protection and avoid directly exposing the information about the state, the variance of the added noise is allowed to be increased for the first time. In addition, the variance of the added noise can be either decaying or constant. By employing a time-varying step-size based on the stochastic approximation method, both the mean-square and almost-sure convergence of the algorithm  are given even with an increasing privacy noise.

\item A guideline for the time-varying step-size and the time-varying variance of the added noise is presented such that the algorithm can achieve the asymptotically unbiased bipartite consensus with the desired accuracy and predefined differential privacy level.

\item The mean-square and almost-sure convergence rate of the algorithm, and the privacy level with different forms of the privacy noises are given. The algorithm's trade-off between the convergence rate and the privacy level is also showed. To the best of our knowledge, it is the first to rigorously characterize both the mean-square and almost-sure convergence rate of distributed consensus in the presence of privacy-preserving, which is in contrast to existing results obtaining the mean-square convergence rate of distributed consensus (e.g., \cite{Nozari2017Differentially}).
\end{itemize}

It is worth noting that this paper's results are significantly different from the literature. A comparison with the state-of-the-art is given as follows. Regarding the noise-perturbation approaches, we remove the requirement that the correlated noises are added to protect the privacy \cite{Mo2017CNPP, He2019CNPPEX}, as well as the requirements of the exponentially decaying \cite{Nozari2017Differentially,Fiore2019DP,Gao2019DP,Zuo2022} and the constant noises in \cite{Liu2020}. Furthermore, we remove the assumption that the adversary doesn't have access to a target agent's communications with all of its neighbors \cite{Ruan2019encryption,Mo2017CNPP,Wang2019,Altafini2020}, and hence, our algorithm protects a more robust privacy of agents regardless of any auxiliary information an adversary may have. Compared with \cite{Mo2017CNPP, He2019CNPPEX,Liu2020,Zuo2022}, both mean-square and almost-sure convergence rate of the algorithm are given. Moreover, we generalize communication topologies from undirected graphs \cite{Mo2017CNPP,He2019CNPPEX,Nozari2017Differentially,Fiore2019DP,Gao2019DP,Liu2020} to a class of signed graphs.

This paper is organized as follows. Section 2 provides the preliminaries and the problem statement. Section 3 introduces the algorithm's convergence and privacy analysis, while Section 4 presents a numerical example. Finally, Section 5 concludes this work.

\textit{Notation.} Denote $\mathbb{R}$, $\mathbb{N}$, $\mathbb{R}_{>0}$ as the sets of the real numbers, nonpositive integers, and positive real numbers, respectively. Let $\mathbb{R}^n$ be the $n$-dimensional real space, and $\mathbb{R}^{n\times m}$ be a set of $n\times m$ real matrices. $I_n$ represents $n\times n$ identity matrix and ${\bf{1}}_n$ is an $n$-dimension column vector with all elements being $1$. The notation ${\rm diag}(b_1,...,b_N)$ denotes the diagonal matrix with diagonal elements $b_1,\ldots, b_N$. For a random variable $X\in \mathbb{R}$, $\mathbb{E}[X]$ and $\text{Var}(X)$ denote the expectation and variance of $X$, respectively. $\text{Lap}(\mu,b)$ denotes the Laplace distribution with mean $\mu$ and scale parameter $b$. $\Gamma(x) = \int\nolimits_0^{+\infty} t^{x-1}e^{-t} \text{d}t$ is the gamma function and  $\Gamma(x,z) = \int\nolimits_z^{+\infty} t^{x-1}e^{-t} \text{d}t$ is the upper incomplete gamma function. For sequences $f(k)$ and $g(k)$ with $k=1,2,\ldots,$  $f(k)=O(g(k))$ means that there exist positive $A$ and $c$ such that $|\frac{f(k)}{g(k)}|\leq A $ for any $k>c$. For any $x\in\mathbb{R}$,  $\text{sgn}(x)$ is the sign function defined as $\text{sgn}(x)=1$ if $x>0$; $-$1 if $x<0$; and 0 if $x=0$. For square matrices $ A_l, \ldots, A_k $, denote $ \prod_{i=l}^k A_i = A_k \cdots A_l $ for $ k\geq l $ and $ \prod_{i=k+1}^{k} A_i = I_n $. For $x \in
\mathbb{R}^n$, $\|x \|_1 = \sum_{i=1}^n |x_i|, \|x\| = \sqrt{
\sum_{i=1}^n x_i^2}$.

\section{Preliminaries and Problem formulation}\label{sec2}

\subsection{Graph theory}
Let $\mathcal{G} = \left( {\mathcal{V},\mathcal{E},\mathcal{A}} \right)$ be an undirected signed graph with a set of agents $\mathcal{V} = \left\{ {1,2, \ldots ,N} \right\}$, a set of edges $\mathcal{E} \in \mathcal{V} \times \mathcal{V}$, and a weighted adjacency matrix $\mathcal{A} = {\left( {{a_{ij}}} \right)_{N \times N}}$. Agent $i$ represents the $i$-th system, and an edge ${e_{ji}}$ in the graph is denoted by the ordered pair agents $\left\{ {j,i} \right\}$. $\left\{ {j,i} \right\} \in \mathcal{E}$ if and only if agent $i$ can obtain the information from agent $j$. For the adjacency matrix $\mathcal{A}$, $a_{ij}\neq0$ if $\left\{ {j,i} \right\} \in \mathcal{E}$, and $a_{ij}=0$, otherwise. Specifically, the interactions between agent $i$ and $j$ is cooperative if $a_{ij}>0$, and competitive if $a_{ij}<0$. We assume there is no self-loop in the graph $\mathcal{G}$, i.e., $a_{ii}=0$. Let $\mathcal{N}_i=\{j|\left\{ {j,i} \right\} \in \mathcal{E}\}$ be a set of agent $i$'s neighbors. The Laplacian matrix $\mathcal{L}=(l_{ij})_{N \times N}$ of graph $\mathcal{G}$ is defined as $l_{ii}=\sum\nolimits_{k = 1,k \ne i}^N {|{a_{ik}}|}$ and  $l_{ij}=-a_{ij}$ if $i\ne j$. We denote $c_{i}=\sum_{j\in\mathcal{N}_i} |a_{ij}|$ as the degree of agent $i$. For a signed graph, we define the greatest degree and the smallest degree as $c_{\max}=\max\{c_i, i\in\mathcal{V}\}$ and $c_{\min}=\min\{c_i, i\in\mathcal{V}\}$. Furthermore,  structural balance is defined as follows.
\begin{definition}\cite{Altafini2013} (Structural balance).
A signed graph $\mathcal{G}$ is structurally balanced if $\mathcal{V}$ can be divided into two disjoint nonempty subsets $\mathcal{V}_{1}$ and $\mathcal{V}_{2}$ (i.e., $\mathcal{V}_{1}\bigcup\mathcal{V}_{2}=\mathcal{V}$ and $\mathcal{V}_{1}\bigcap\mathcal{V}_{2}=\emptyset$) such that $a_{ij}\geq0$ for $\forall i, j\in \mathcal{V}_{h} (h\in\{1,2\})$, and $a_{ij}\leq0$ for $\forall i\in \mathcal{V}_{h}, j\in \mathcal{V}_{q}, h\neq q,(h, q\in\{1,2\})$.
\end{definition}

\begin{assumption} \label{Assumption_Graph}
The signed graph $\mathcal{G}$ is connected and structurally balanced.
\end{assumption}

\begin{lemmax}\cite{Altafini2013}\label{lemma1}
If Assumption \ref{Assumption_Graph} holds, then

\begin{enumerate}
\item A diagonal matrix $S={\rm{diag}}(s_{1}, s_{2},\ldots, s_{n})$ exists, such that $S\mathcal{A}S$ has all nonnegative elements, where $s_{i}\in\{1, -1\}$, for all $i\in\mathcal{V}$.
\item The Laplacian matrix  associated with the corresponding unsigned graph $\mathcal{L}_S=S\mathcal{L}S$ is positive semi-definite.
\item The eigenvalue $\lambda_{k}(\mathcal{L}), k=1, 2,\ldots, n$ of the Laplacian matrix $\mathcal{L}$ satisfies $0=\lambda_{1}(\mathcal{L})<\lambda_{2}(\mathcal{L})\leq \ldots\leq \lambda_{n}(\mathcal{L})$.
\end{enumerate}
\end{lemmax}
\begin{remark}\label{remark:LS}
Assumption \ref{Assumption_Graph} describes the connectivity of the signed graph and is standard in the bipartite consensus \cite{Altafini2013}. Under Assumption \ref{Assumption_Graph}, we have ${\bf 1}_N^TS\mathcal{L}=0$. In particular, if $S=I$, then the problem studied in this paper is reduced to the traditional consensus problem \cite{Huang2012DP,Nozari2017Differentially}.
\end{remark}
\subsection{Problem formulation}
Consider a set of $N$ agents coupled by an undirected signed graph $\mathcal{G}$. The dynamics of the $i$-th agent are as follows
\begin{eqnarray} \label{system}
\begin{array}{l}
x_i(k+1)=x_i(k)+u_i(k),\\
\end{array}
\end{eqnarray}
where $x_i(k)\in\mathbb{R}$ is the state of agent $i$ with initial value $x_i(0)$, and ${u_i}(k) \in {\mathbb{R}}$ is the control input. To achieve the bipartite consensus of the system (\ref{system}), \cite{Liu2017} designs the following distributed controller:
\begin{eqnarray*}
\begin{array}{l}
u_i(k)=-\sum\limits_{j\in\mathcal{N}_{i}}|a_{ij}|(x_i(k)-\text{sgn}(a_{ij})x_{j}(k)),
\end{array}
\end{eqnarray*}
where $x_j(k)$ is the information that agent $i$ receives from its neighbors $j$.

In distributed bipartite consensus, {\it eavesdroppers} or {\it honest-but-curious (semi-honest) agents} may exist in the network. Note that the so-called {\it honest-but-curious agents} might collude and attempt to deduce information about the initial state values of the other honest agents from the information they receive. {\it Eavesdroppers} are external adversaries who steal information through wiretapping all communication channels and intercepting exchanged information between agents. An {\it honest-but-curious agent} $i$ has access to the internal state $x_{i}$, which is unavailable to external {\it eavesdroppers}. However, an {\it eavesdropper} has access to all shared information in the network, whereas an {\it honest-but-curious agent} can only access shared information that is destined to it. These two attacker types are collectively called {\it passive attackers}. If the network has {\it passive attackers}, delivering $\{x_i(k)|k\ge 0\}$ directly for each agent may leak its privacy, including the state $x_i(k)$ and the initial opinion or belief $x_i(0)$. Therefore, direct communication of intermediate results in the above controller can lead to severe privacy leakage of each agent's sensitive information. It is imperative to provide a theoretical privacy guarantee on each agent's sensitive information.

\subsection{Differential privacy}
A mechanism $\mathcal{M}(\cdot)$ is a stochastic map from a private dataset $D$ to an observation $O$. In this paper, we focus on protecting the initial states of each agent against passive attackers. Thus, the private dataset is $D=\{x_i(0), i\in \mathcal{V}\}$, and the observation is $O = \{y_i(k), i\in \mathcal{V} \}_{k=0}^{T-1}$.  Then, we introduce the $\epsilon$-differential privacy for the private dataset.
\begin{definition}\cite{Nozari2017Differentially} For any given $\delta\in\mathbb{R}$, the initial states $D=\{x_i(0), i\in \mathcal{V}\}$ and $D' = \{x'_i(0), i\in \mathcal{V}\}$ are $\delta$-adjacent if there exists $i_{0}\in\mathcal{V}$, such that
\begin{eqnarray*}
|x_{i}(0)-x'_{i}(0)|
\leq
\left\{ \begin{array}{cc}
\delta & i=i_{0} \\
 0 & i\neq i_{0}
 \end{array}
\right.
\end{eqnarray*}
\end{definition}
Based on the above definition, inspired by \cite{Ny2014DP,Nozari2017Differentially}, a definition of differential privacy is given for the differentially private bipartite consensus as follows.
\begin{definition} (Differential privacy).
Given $\epsilon>0$ and $\delta>0$, for any two $\delta$-adjacent initial states $D=\{x_i(0), i\in \mathcal{V}\}$, $D' = \{x'_i(0), i\in \mathcal{V}\}$ and an observation set $\mathcal{O}\subseteq(\mathbb{R}^{N})^{\mathbb{N}}$, the mechanism $\mathcal{M}(\cdot)$ is $\epsilon$-differentially private if $\mathbb{P}\{\mathcal{M}(D)\in \mathcal{O}\}\leq e^{\epsilon} \mathbb{P}\{\mathcal{M}(D')\in \mathcal{O}\}$ holds, where $\epsilon$ is the privacy level.
\end{definition}

Next, the asymptotically unbiased mean-square bipartite consensus, almost-sure bipartite consensus, and $(s,r)$-accuracy are defined as follows, respectively.
\begin{definition} \label{DefinitionConsensus}
(Mean-square bipartite consensus).
The system (\ref{system}) is said to achieve the mean-square bipartite consensus if for any given initial state, there exists a random variable $x^\star$ with $\mathbb{E}[x^{\star}]^{2}<\infty$, such that
$\lim\limits_{k\rightarrow\infty}\mathbb{E}[x_{i}(k)-s_{i}x^{\star}]^{2} = 0, s_{i}\in\{1,-1\}, \forall~i\in\mathcal{V}.$
\end{definition}

\begin{definition}
(Almost-sure bipartite consensus).
The system (\ref{system}) is said to achieve the almost-sure bipartite consensus if for any given initial state, there exists a random variable $x^{\star}\in\mathbb{R}$ with $\mathbb{E}[x^{\star}]^{2}<\infty$, such that
$\lim\limits_{k\rightarrow\infty}x_{i}(k)=s_{i}x^{\star}, s_{i}\in\{1,-1\}, \forall ~i\in\mathcal{V}.$
\end{definition}

\begin{definition}(Accuracy).
For $s\in [0,1]$ and $r\in\mathbb{R}_{\ge 0}$, the system (\ref{system}) is said to achieve an $(s,r)$-accuracy, if for any given initial state, there exists a random variable $x^\star$, such that $\mathbb{P}\{|x^\star-\frac{{\bf 1_{N}^T}Sx(0)}{N}|\leq r\}\ge 1-s$, where $\mathbb{E}x^\star=\frac{{\bf 1_{N}^T}Sx(0)}{N}$, $\rm{Var}$$(x^\star)$ is bounded.
\end{definition}

{\bf Problems of interest}: In this paper, the following questions are answered:

\begin{itemize}
\item How to design a more general privacy noise form for distributed bipartite consensus to achieve a better privacy protection with guaranteed convergence?

\item What is the mean-square and almost-sure convergence rate under the influence of privacy noises?

\item How to design a distributed protocol $u_i(k)$ and privacy noises for the desired accuracy $(s^\star, r^\star)$ and predefined differential privacy level $\epsilon^\star$?
\end{itemize}
\section{Main result}
This section first presents the convergence analysis, ensuring that the asymptotically unbiased mean-square and almost-sure bipartite consensus are achieved under certain conditions. In addition to the mean-square and almost-sure convergence rate, the privacy analysis is also given by introducing the definition of sensitivity on a private dataset. Finally, the trade-off between convergence rate and privacy level is discussed.
\subsection{Algorithm}
This subsection introduces two steps of the proposed algorithm: {\bf information transmission} and {\bf state update}.

{\bf Information transmission}: Each agent $i$ sends to its neighbors the following information instead of the original information $x_i(k)$.
 \begin{eqnarray} \label{NoiseAdd}
y_i(k)= x_i(k)+\omega_i(k),  i\in\mathcal{V}, k\in \mathbb{N}_{>0}
\end{eqnarray}
where the random variable $\omega_i(k)\in\mathbb{R}$ follows Laplace distribution with a variance of $2b^2(k)$, i.e., $\omega_{i}(k) \sim \text{Lap}(0, b(k)).$ Moreover, $\omega_i(k)$ and $\omega_j(k)$, $\forall i,j\in \mathcal{V}$, $i\neq j$ are mutually independent.

{\bf State update}: Each agent $i$ receives $y_j(k)$ from its neighbor $j$  and updates its own state by using the following privacy-preserving distributed controller:
\begin{eqnarray} \label{controller}
\begin{array}{l}
u_i(k)=-\alpha(k)\sum\limits_{j\in\mathcal{N}_{i}}|a_{ij}|(x_i(k)-\text{sgn}(a_{ij})y_{j}(k)),
\end{array}
\end{eqnarray}
where $\alpha(k)$ is a positive time-varying step-size. Then, each agent $i$ updates its own state as follows.
\begin{align}\label{closedsystem}
x_i(k+1)\!\!=\!\!x_i(k)\!\!-\!\!\alpha(k)\sum\limits_{j\in\mathcal{N}_{i}}|a_{ij}|(x_i(k)-\text{sgn}(a_{ij})y_{j}(k)).
\end{align}

Set
\begin{align*}
x(k)&= \begin{bmatrix}
x_1(k)& x_2(k)& \ldots& x_N(k)
\end{bmatrix}^T,\\
y(k) &= \begin{bmatrix}
y_1(k)& y_2(k)& \ldots& y_N(k)
\end{bmatrix}^T,\\
\omega(k) &= \begin{bmatrix}
\omega_1(k)& \omega_2(k)& \ldots& \omega_N(k)
\end{bmatrix}^T.
\end{align*}
Then, the equation (\ref{closedsystem}) can be rewritten in a compact form as follows:
\begin{align}\label{closedsystem1}
x(k+1)=\left(I-\alpha(k)\mathcal{L}\right)x(k)+\alpha(k)\mathcal{A}\omega(k).
\end{align}
\begin{remark}
In order to achieve privacy protection, we add a noise to agent $i$'s state before transmitting it to its neighbors. Different from the existing literature, the privacy noise added in (\ref{NoiseAdd}) is more generic. Specifically, the privacy noise used in this paper is random with a variance of ${2b^2(k)}$, which is not required to decay to zero as in the existing literature. Therefore, the state's information is not directly inferred with time. However, this brings convergence difficulties with the corresponding privacy analysis described next. Moreover, it is worth noting that we employ a time-varying step-size $\alpha(k)$, making the controller more flexible than that utilizing a constant step-size. Note that if $\alpha(k)$ is set to constant, as in current literature, the above closed-loop system cannot achieve asymptotically unbiased convergence because of the influence of privacy noises (\ref{NoiseAdd}) that will not decay to zero. To this end, we apply the stochastic approximation method to design a time-varying step-size.
\end{remark}

\subsection{Convergence analysis}
This subsection first proves that the algorithm can achieve asymptotically unbiased mean-square and almost-sure bipartite consensus. Then, we provide a method to design the step-sizes $\alpha(k)$ and noise parameters $b(k)$ to ensure the $(s^\star, r^\star)$-accuracy.

\begin{lemmax}\cite{Polyak1987} \label{Lemma_Appendix}
Let $\{u(k), k\in\mathbb{N}\}$, $\{\alpha(k), k\in\mathbb{N}\}$ and  $\{q(k), k\in\mathbb{N}\}$ be real sequences, satisfying $0<q(k)\le 1$, $\alpha (k)\ge 0$, $\sum_{k=0}^{\infty}q(k)=\infty$, $\lim_{k\to\infty} \alpha(k)/q(k)=0$  and
$u(k+1)=(1-q(k))u(k)+\alpha(k).$ Then $\lim \sup_{k\to \infty}u(k)\le 0$. In particular, if $u(k)\ge 0$, $k\in\mathbb{N}$, then $\lim_{k\to\infty}u(k)=0$.
\end{lemmax}

For the step-size $\alpha(k)$ and the noise parameter $b(k)$, we give the following assumption.
\begin{assumption}\label{Assumption step}
The step-size $\alpha(k)$ and the noise parameter $b(k)$ satisfy one of the following conditions:
\begin{item}
\item \hspace*{0.4cm}\textbf{a)} $\sum\limits_{k=0}^\infty\alpha(k)=\infty$, $\lim\limits_{k\rightarrow\infty}\alpha(k)=0$,
    $\sum\limits_{k=0}^\infty\alpha^2(k)b^2(k)<\infty$;

\item \hspace*{0.4cm}\textbf{b)} $\sum\limits_{k=0}^\infty\alpha(k)=\!\infty$, $\sum\limits_{k=0}^\infty\alpha^2(k)<\infty$, $\sum\limits_{k=0}^\infty\alpha^2(k)b^2(k)<\infty$.
\end{item}
\end{assumption}
\begin{remark}
Assumption \ref{Assumption step} \textbf{a)} is weaker than Assumption
\ref{Assumption step}~\textbf{b)}. For example, Assumption \ref{Assumption step} \textbf{a)} and \textbf{b)} are satisfied for a step-size in the form of $\alpha(k)=(k+1)^{-\beta}$, $\beta\in(1/2,1]$, and the noise parameter in the form of $b(k)=(k+1)^{\gamma}, \gamma<\beta-1/2$.
However, if $\alpha(k)=(k+1)^{-\beta}$, $\beta\in(0,1/2]$, and $b(k)=(k+1)^{\gamma}, \gamma<\beta-1/2$, then
Assumption~\ref{Assumption step}~\textbf{a)} holds, but Assumption \ref{Assumption step} \textbf{b)} fails.
\end{remark}
\begin{theorem}\label{Theorem_Convergence}
If Assumptions \ref{Assumption_Graph} and \ref{Assumption step}~\textbf{a)} hold, then the mean-square bipartite consensus of the algorithm can be achieved for all $i\in\mathcal{V}$.
\end{theorem}
\textit{Proof}:
Let $z(k)=Sx(k)$ and $\mathcal{L}_{S}=S\mathcal{L}S$. Then, from (\ref{closedsystem1}) and noting $S^{-1}=S$ it follows that
\begin{align}\label{consensusstate}
z(k+1)=&\left(I-\alpha(k)S\mathcal{L}S^{-1}\right)z(k)+\alpha(k)S\mathcal{A}\omega(k)\nonumber\\
=&\left(I-\alpha(k)S\mathcal{L}S\right)z(k)+\alpha(k)S\mathcal{A}\omega(k)\nonumber\\
=&\left(I-\alpha(k)\mathcal{L}_{S}\right)z(k)+\alpha(k)S\mathcal{A}\omega(k)
\end{align}
Let $J=(1/N)\mathbf{1}\mathbf{1}^{T}$, $\delta(k)=(I_{N}-J)z(k)$, $V(k)=\|\delta(k)\|^2$ and note $\mathbf{1}^{T}\mathcal{L}_{S}=0$ and $J\mathcal{L}_{S}=0$. Then, from (\ref{consensusstate}), we have
\begin{eqnarray*}
\delta(k+1)
&=& (I_{N}-J)z(k+1)\cr
&=&z(k)-\alpha(k)\mathcal{L}_{S}z(k)+ \alpha(k)S\mathcal{A}\omega(k)\cr
\!\!&& -Jz(k)-\alpha(k)JS\mathcal{A}\omega(k)\cr
\!\!&=&\delta(k)-\alpha(k)\mathcal{L}_{S}z(k)+\alpha(k)(I_{N}-J)S\mathcal{A}\omega(k)\cr
\!\!&=&\!\!\big[I_{N}-\alpha(k)\mathcal{L}_{S}\big]\delta(k)+\alpha(k)(I_{N}-J)S\mathcal{A}\omega(k)
\end{eqnarray*}
Thus,
\begin{eqnarray}\label{20}
\!\!&\!\!\!\!&\!\!V(k+1)\cr
\!\!&\!\!=&V(k)\!-\!\alpha(k)\delta^{T}(k)(\mathcal{L}_{S}\!\!+\!\!\mathcal{L}^{T}_{S})\delta(k)\!\!+\!\! \alpha^{2}(k)\delta^{T}(k)\mathcal{L}^{T}_{S}\mathcal{L}_{S}\delta(k)\cr
\!\!&\!\ \!\!&\!\! +2 \alpha(k) \delta^{T}(k)\big[I_{N}-\alpha(k)\mathcal{L}^{T}_{S}\big](I_{N}-J)S\mathcal{A}\omega(k)\cr
\!\!&\!\!\!\!&\!\!+\alpha^{2}(k)w^{T}(k)\mathcal{A}^{T}S^{T}(I_{N}-J)^{T}(I_{N}-J)S\mathcal{A}\omega(k)
\end{eqnarray}
Since $\mathbf{1}\delta(k)=0 $ and $\lambda_{2}(\mathcal{L}_{S})=\lambda_{2}(\mathcal{L})$, from Theorem 2.1 in \cite{Li2010CommunicationNoise}, we have $\delta^{T}(k)\mathcal{L}_{S}\delta(k) \geq \lambda_{2}(\mathcal{L})V(k)$, which together with \eqref{20} and the symmetry of $ \mathcal{L}_S $ implies
\begin{align*}
&V(k+1)\nonumber\\
\leq& \left( 1 - 2\alpha(k)\lambda_{2}(\mathcal{L})+ \alpha^{2}(k)\|\mathcal{L}_{S}\|^{2}\right) V(k)\nonumber\\
	&+2\alpha(k) \delta^{T}(k)\big[I_{N}-\alpha(k)\mathcal{L}^{T}_{S}\big](I_{N}-J)S\mathcal{A}\omega(k)\nonumber\\
	&+\alpha^{2}(k)\omega^{T}(k)\mathcal{A}^{T}S^{T}(I_{N}-J)^T(I_{N}-J)S\mathcal{A}\omega(k).
\end{align*}
Define $\sigma$-algebra $\mathcal{F}^{\omega}_k=\sigma\{\omega(0),\omega(1),\omega(2),\ldots, \omega(k)\}$. Note that $\omega(k)$ is the zero-mean noise. Then, from Assumption \ref{Assumption_Graph} and taking the conditional expectation with respect to $\mathcal{F}^{\omega}_k$ on both sides of the above equations, on can get
\begin{eqnarray}\label{LyapunovUpdateE1}
&&\mathbb{E}\left[V(k+1)|\mathcal{F}^{\omega}_k\right]\cr
\!\!&\!\!\leq&(1-2\alpha(k)\lambda_2(\mathcal{L})+ \alpha^{2}(k)\|\mathcal{L}_{S}\|^{2})V(k)\cr
\!\!&\!\!\!\!&\!\!+\alpha^{2}(k)\|I_{N}-J\|^{2}\|S\|^{2}\|\mathcal{A}\|^{2}\mathbb{E}\|\omega(k)\|^{2}\cr
\!\!&\!\!=&(1-2\alpha(k)\lambda_2(\mathcal{L})+ \alpha^{2}(k)\|\mathcal{L}_{S}\|^{2})V(k)\cr
\!\!&\!\!\!\!&\!\!+2\alpha^{2}(k)b^{2}(k)N\|I_{N}-J\|^{2}\|S\|^{2}\|\mathcal{A}\|^{2}.
\end{eqnarray}
Note that
\begin{eqnarray*}
\mathbb{E}\left[\mathbb{E}\left[V(k+1)|\mathcal{F}^{\omega}_k \right]\right]=\mathbb{E}\left[V(k+1)\right].
\end{eqnarray*}
Then, taking mathematical expectation on both sides of (\ref{LyapunovUpdateE1}), we obtain
\begin{eqnarray}\label{22}
\mathbb{E}\left[V(k+1)\right]
\leq&(1-2\alpha(k)\lambda_2(\mathcal{L})+\alpha^{2}(k)\|\mathcal{L}_{S}\|^{2})V(k)\cr
&+2\alpha^{2}(k)b^{2}(k)N\|I_{N}-J\|^{2}\|S\|^{2}\|\mathcal{A}\|^{2}.
\end{eqnarray}
Note that $\lim_{k\to\infty}\alpha(k)=0$. Then, there exists $M$ such that $\alpha(k) \le \min\left\{ \frac{\lambda_2(\mathcal{L})}{\|\mathcal{L}_{S}\|^2}, \frac{1}{2\lambda_2(\mathcal{L})} \right\}$ for all $k>M$, which implies $\alpha(k)\|\mathcal{L}_{S}\|^2\leq\lambda_2(\mathcal{L})$ and $2\alpha(k)\lambda_2(\mathcal{L})\leq1$. Thus, by Assumption \ref{Assumption step}~\textbf{a)} we have
\begin{eqnarray*}
&&\!\!\!0\leq1-2\alpha(k)\lambda_2(\mathcal{L})+\alpha^2(k)\|\mathcal{L}_{S}\|^2 <1, \forall k\geq M, \\
&&\!\!\!\sum_{k=0}^{\infty} (2\alpha(k)\lambda_2(\mathcal{L})-\alpha^2(k)\|\mathcal{L}_{S}\|^2)\ge \lambda_2(\mathcal{L})\sum_{k=0}^{\infty}\alpha(k) =\infty, \\
&&\!\!\!\lim_{k \to \infty} \frac{2\alpha^2(k)b^2(k)N\|I_{N}-J\|^{2}\|S\|^{2}\|\mathcal{A}\|^{2}}
{2\alpha(k)\lambda_2(\mathcal{L})-\alpha^2(k)\|\mathcal{L}_{S}\|^2} =0.
\end{eqnarray*}
Therefore, by Lemma \ref{Lemma_Appendix}, we have $\lim_{k\to\infty} \mathbb{E}[V(k)] = 0$. This completes the proof.   $\hfill\Box$

\begin{theorem} \label{BiasedConsensus}
If Assumptions \ref{Assumption_Graph} and \ref{Assumption step} \textbf{a)} hold, then the algorithm converges in mean-square for all $i\in\mathcal{V}$, i.e., $\lim_{k\to\infty} \mathbb{E}\left[x_i(k)-s_{i}x^\star \right]^2=0$, where $x^\star$ is a random variable, satisfying $\mathbb{E}[x^\star] = \frac{1}{N} {{\bf{1}}_{N}^T} Sx(0)$ and ${\rm{Var}}$$[x^\star]=\frac{\sum_{i\in \mathcal{V}} c_i^2}{N^{2}}\sum_{j=0}^\infty \alpha^2(j)b^2(j)$.
\end{theorem}
\textit{Proof}:
Since the graph is structurally balanced, from Lemma~\ref{lemma1} it follows that $\textbf{1}_N\mathcal{L}_{S}=0$, and
\begin{align*}
{\bf{1}}_{N}^T z(k)
&= ({\bf{1}}_{N}^T(I_N-\alpha(k-1)\mathcal{L}_{S}))z(k-1)\\
&~~~+ \alpha(k-1) ({{\bf{1}}_{N}^T}S\mathcal{A})\omega(k-1)\\
&={\bf{1}}_{N}^T z(k-1) + \alpha(k-1) ({{\bf{1}}_{N}^T}S\mathcal{A})\omega(k-1).
\end{align*}
By iteration, we have
\begin{align}\label{eq:sum ave z}
{{\bf{1}}_{N}^T}z(k)=\sum_{i\in \mathcal{V}} z_{i}(0)+\sum_{j = 1}^{k}\alpha(j-1) ({{\bf{1}}_{N}^T} S\mathcal{A})\omega(j-1),
\end{align}
which immediately follows that
\begin{align*}
\lim_{k\to \infty} {{\bf{1}}_{N}^T}z(k) =  \sum_{i\in \mathcal{V}}  z_{i}(0)+\sum_{j =1}^{\infty} \sum_{i\in \mathcal{V}}\alpha(j-1) s_i c_i\omega_{i}(j-1).
\end{align*}
By Theorem \ref{Theorem_Convergence}, set $x^\star\!\!=\!\!\frac{1}{N} \sum_{i\in \mathcal{V}} z_i(0)\!+\!\frac{1}{N}\sum_{j =1}^{\infty}\sum_{i\in \mathcal{V}} \alpha(j-1)s_ic_i\omega_i(j-1)$. Then, we have
\begin{align*}
&\lim_{k\to\infty} \sqrt{\mathbb{E}\left[s_{i}x_i(k)-x^\star\right]^2}\nonumber\\
\le & \lim_{k\to\infty} \sqrt{\mathbb{E}\left[s_{i}x_i(k)-\frac{1}{N}\textbf{1}_{N}^Tz(k)\right]^2}  \nonumber\\
&+ \lim_{k\to\infty} \sqrt{\mathbb{E}\left[\frac{1}{N}\textbf{1}_{N}^Tz(k)-x^\star\right]^2}=0.
\end{align*}
By the fact that $\omega_i(k-1)$ are i.i.d. for all $i\in \mathcal{V}$, $k\in \mathbb{N}_{> 0}$, it is obtained that
\begin{align*}
\mathbb{E}x^\star &= \mathbb{E}\left[\frac{1}{N} \sum_{i\in \mathcal{V}}  z_i(0) + \frac{1}{N}\sum_{j =0}^{\infty} \sum_{i\in \mathcal{V}} \alpha(j) s_i c_i\omega_i(j-1)\right] \nonumber\\
&= \frac{1}{N} \sum_{i\in \mathcal{V}}z_i(0) =\frac{1}{N} \sum_{i\in \mathcal{V}} s_{i}x_i(0),
\end{align*}
and
\begin{align} \label{Formula_Varystar}
{\text{Var}}\left(x^\star\right)&= \frac{1}{N^2}\sum_{j=0}^\infty \sum_{i\in \mathcal{V}} \alpha^2(j) \mathbb{E}\left[s_ic_i \omega_i(j)\right]^{2}\nonumber\\
&= \frac{1}{N^{2}}\sum_{j=0}^\infty \sum_{i\in \mathcal{V}} \alpha^2(j) c_i^{2}\mathbb{E}\left[ \omega_i(j)\right]^{2} \nonumber\\
&=\frac{2\sum_{i\in \mathcal{V}} c_i^2}{N^{2}}\sum_{j=0}^\infty \alpha^2(j)b^2(j).
\end{align}
Since $\sum_{j=0}^\infty \alpha^2(j)b^2(j)<\infty$, ${\text{Var}}\left(x^\star\right)$ is bounded. This completes the proof.

Note that almost-sure convergence properties are important, because they represent what
happen to individual trajectories of the stochastic iterations, which are instantiations of the algorithm actually used in practice. From the following theorem it follows that under Assumptions \ref{Assumption_Graph} and \ref{Assumption step} \textbf{b)}, for a class of privacy noises, the almost-sure bipartite consensus of the algorithm is achieved as well.
\begin{theorem}\label{Theorem_Convergence3}
If Assumptions \ref{Assumption_Graph} and \ref{Assumption step} \textbf{b)} hold, then the algorithm converges almost-surely for all $i\in\mathcal{V}$, i.e. $\lim\limits_{k\rightarrow\infty}x_{i}(k)=s_{i}x^{\star}, a.s.,$ where $s_{i}\in\{1,-1\}$.
\end{theorem}
\textit{Proof}:
From (\ref{LyapunovUpdateE1}) it follows that
\begin{eqnarray*}
&&\mathbb{E}\left[V(k+1)|\mathcal{F}^{\omega}_k\right]\nonumber\\
&\le &
\left(1+\alpha^2(k)\|\mathcal{L}_{S}\|^2\right) V(k)-2\alpha(k)\lambda_2(\mathcal{L})V(k)\nonumber\\
&~&+\alpha^2(k)\|I_{N}-J\|^{2}\|S\|^{2}\|\mathcal{A}\|^{2}\mathbb{E}\|\omega(k)\|^{2}\nonumber\\
&\le &\left(1+\alpha^2(k)\|\mathcal{L}_{S}\|^2\right) V(k)-2\alpha(k)\lambda_2(\mathcal{L})V(k)\nonumber\\
&~&+2\alpha^2(k)b^2(k)N\|I_{N}-J\|^{2}\|S\|^{2}\|\mathcal{A}\|^{2}.
\end{eqnarray*}
Notice that $\sum\limits_{k=0}^{\infty}\alpha^2(k)b^2(k)<\infty$ a.s. and
$\sum\limits_{k=0}^{\infty}\alpha^2(k)<\infty$ a.s.. Then, based on the nonnegative supermartingale convergence theorem \cite{Goodwin1984}, $V(k)$ converges to 0 almost-surely, and
\begin{align*}
\sum_{k=0}^{\infty}\alpha(k)V(k)<\infty ~~a.s.
\end{align*}
By Assumption \ref{Assumption step} \textbf{b)}, we have
\begin{align*}
&\mathbb{E} \left\|\sum_{j = 1}^{k}\alpha(j-1) ({{\bf{1}}_{N}^T} S\mathcal{A})\omega(j-1)\right\|^2\\
\le &\left\| {{\bf{1}}_{N}^T} S\mathcal{A} \right\|^2  \sum_{j=1}^{k} \mathbb{E} \left\| \alpha(j-1)\omega(j-1)\right\|^2 \\
= & 2\left\| {{\bf{1}}_{N}^T} S\mathcal{A} \right\|^2\sum_{j=1}^{k}  \alpha^2(j-1) b^2(j-1)<\infty.
\end{align*}
This together with (\ref{eq:sum ave z}) and Theorem 7.6.10 of \cite{Ash1972} implies that
 $ \frac{1}{N}\sum_{i=1}^N z_i(k) $ converges to $x^\star$ almost-surely. Since $V(k)$ converges to 0 almost-surely, we have $z_i(k) $ converges to $ \frac{1}{N}\sum_{i=1}^N z_i(k) $ almost-surely.  This proves the theorem.   $\hfill\Box$

\begin{remark}
Theorems \ref{Theorem_Convergence}-\ref{Theorem_Convergence3} show that the state of each agent converges to the average of the states in mean-square and almost-sure. Both theorems also hold for the decaying noises \cite{Nozari2017Differentially,Zuo2022} in the form of $\omega_{i}(k) \sim \rm{Lap}$$(0, c_iq_i^k),$ where $\omega_i(k)$ and $\omega_j(k)$ are mutually independent of $c_i>0$, $0<q_i<1$. This is because, under the above decaying noises, it still holds that $\{\omega(k),\bar{\mathcal{F}}^{\omega}_k\}_{k=1}^{\infty}$ is a martingale difference sequences and $b^2(k) \triangleq\sup_{k\in\mathbb{N}_{>0}} \mathbb{E}\left[\|\omega(k)\|^2\right]$ is bounded. Therefore, by using the stochastic approximation method, we relax the selection of the privacy noises, comparing with the existing literature. However, this method rises another problem, namely how to design the  step-size $\alpha(k)$ and the noise parameters $b(k)$ to satisfy the $(s^\star, r^\star)$-accuracy and $\epsilon^\star$-differential privacy requirements.
\end{remark}

The following theorem provides a way to design the step-size $\alpha(k)$ and the noise parameters $b(k)$ to ensure the $(s^\star, r^\star)$-accuracy.

\begin{theorem}  \label{Preaccuracy}
Suppose Assumption \ref{Assumption_Graph} holds. Given a pair of parameters $(s^\star, r^\star)$, if we set $\alpha(k)=\frac{a_1}{(k+a_2)^{\beta}}$, $\beta\in (0,1]$, $a_1, a_2>0$, and $b(k)=\underline{b}(k+a_2)^{\gamma}$, $\gamma<\beta-1/2$, $\underline{b}>0$, such that
\begin{eqnarray}  \label{ConditionAccuracy}
 \frac{a_1^2\underline{b}^2a_2^{2\gamma-2\beta+1}}{2\beta-2\gamma-1} +\frac{a_1^2\underline{b}^2a_2^{2\gamma}}{a_2^{2\beta}} \le \frac{s^\star (r^\star)^2N^{2}}{2\sum_{i\in \mathcal{V}} c_i^2},
 \end{eqnarray}
then the $(s^\star, r^\star)$-accuracy is ensured, i.e.,
\begin{eqnarray*}
\mathbb{P} \left\{|x^\star-\mathbb{E}x^\star|<r^\star\right\}\ge 1-s^\star.
\end{eqnarray*}
\end{theorem}
\textit{Proof}:
From the Chebyshev's inequality \cite{Ash1972} it follows that
\begin{eqnarray*}
\mathbb{P} \left\{\frac{(x^\star-\mathbb{E}x^\star)^2}{{\text{Var}}(x^\star)}<\epsilon \right\}\ge 1-\frac{1}{\epsilon}.
\end{eqnarray*}
Taking (\ref{Formula_Varystar}) into the above inequality yields
\begin{eqnarray*}
\mathbb{P} \left\{|x^\star-\mathbb{E}x^\star| < \sqrt{\epsilon\kappa} \right\}\ge 1-\frac{1}{\epsilon},
\end{eqnarray*}
where $\kappa = \frac{\sum_{i\in \mathcal{V}} c_i^2}{N^{2}}\sum_{k=0}^\infty \alpha^2(k)b^2(k)$. Set $r=\sqrt{\epsilon \kappa}$. Then, $\epsilon=\frac{r^2}{\kappa}$ and
\begin{eqnarray*}
\mathbb{P} \left\{|x^\star-\mathbb{E}x^\star|<r\right\}\ge 1-\frac{\kappa}{r^2}.
\end{eqnarray*}
Therefore, the $(s,r)$-accuracy is achieved with $s\!\!=\!\!\frac{2\sum_{i\in \mathcal{V}} c_i^2}{N^{2}r^2}\sum_{k=0}^\infty \alpha^2(k)b^2(k)$.

Clearly, as long as $\sum_{k=0}^\infty \alpha^2(k) b^2(k) \le \frac{s^\star (r^\star)^2N^{2}}{2\sum_{i\in \mathcal{V}} c_i^2},$ the $(s^\star, r^\star)$-accuracy is ensured. By the fact that $f(x)=\frac{a_1\underline{b}(x+a_{2})^{\gamma}}{(x+a_2)^{\beta}}$ with $\beta\in (0,1]$, $a_1$, $a_2$, $\underline{b}>0$, is a strictly decreasing function of $x>0$, for $k\in \mathbb{N}_{> 0}$, we have $\left(\frac{a_1\underline{b}(k+a_{2})^{\gamma}}{(k+a_2)^{\beta}}\right)^2 \le \int_{{k-1}}^k  \left( \frac{a_1\underline{b}(x+a_{2})^{\gamma}}{(x+a_2)^{\beta}}\right)^2 \text{d}x$, and thus,
\begin{eqnarray*}
&&\sum_{k=0}^\infty \alpha^2(k) b^2(k) \nonumber\\
&=& \frac{a_1^2\underline{b}^2a_2^{2\gamma}}{a_2^{2\beta}}+\sum_{k=1}^\infty \left( \frac{a_1\underline{b}(k+a_{2})^{\gamma}}{(k+a_2)^{\beta}}\right)^2\nonumber\\
&\le &  \frac{a_1^2\underline{b}^2a_2^{2\gamma}}{a_2^{2\beta}}+ \int_{0}^\infty  \left( \frac{a_1\underline{b}(x+a_{2})^{\gamma}}{(x+a_2)^{\beta}}\right)^2 \text{d}x\nonumber\\
&\le &   \left. \left(\frac{a_1^2\underline{b}^2 (x+a_2)^{2\gamma-2\beta+1}}{2\gamma-2\beta+1}\right) \right|_0^{\infty} +\frac{a_1^2\underline{b}^2a_2^{2\gamma}}{a_2^{2\beta}}\nonumber\\
&\le & 0-\frac{a_1^2\underline{b}^2a_2^{2\gamma-2\beta+1}}{2\gamma-2\beta +1} +\frac{a_1^2\underline{b}^2a_2^{2\gamma}}{a_2^{2\beta}} \le \frac{s^\star (r^\star)^2N^{2} }{2\sum_{i\in \mathcal{V}} c_i^2}.
\end{eqnarray*}
This completes the proof. $\hfill\Box$

Under the time-varying noises, the predefined accuracy can be ensured by selecting a proper step-size $\alpha(k)$ and noise parameter $b(k)$, $k\in\mathbb{N}_{\ge 0}$. Besides, we can enhance the accuracy by minimizing the term $\sum_{k=0}^\infty \alpha^2(k)b^{2}(k)$.

\subsection{Convergence rate}
In this subsection, we first provide two lemmas, and then, give the mean-square and almost-sure convergence rate of the algorithm with $\alpha(k)=\frac{\alpha_k}{(k+a_{2})^\beta}$ and $b(k)=O(k^\gamma)$.

\begin{lemmax}\label{lemma:prod_est}
For $ 0<\beta\leq 1,\ \alpha,\ l>0,\ k_0\geq0$, we have
	\begin{align}\label{ineq:prod}
&\prod_{i=l}^k \left(1-\frac{\alpha}{(i+k_0)^\beta}\right)\nonumber\\
\leq& \!\!
\begin{cases}
\left( \frac{l+k_0}{k+k_0} \right)^\alpha, & \beta =1;\\
\exp\left( \frac{\alpha}{1-\beta}\left((l\!+\!k_0)^{1-\beta}\!-\!(k\!+\!k_0\!+\!1)^{1-\beta}\right) \right), & \!\!\!\beta\in(0,1).
		\end{cases}
	\end{align}
If we further assume that $\beta>1/2$, then for any $ \gamma>0 $, we have
	\begin{align}\label{est:prod}
&\prod_{i=l}^k \left(1-\frac{\alpha}{(i+k_0)^\beta}+\frac{\gamma}{(i+k_0)^{2\beta}}\right)\nonumber\\
\!\!=\!\!&\begin{cases}
			\!O\!\left(\left( \frac{l+k_0}{k+k_0} \right)^\alpha\right), & \beta =1;\\
			\!O\!\left(\!\exp\!\left(\! \frac{\alpha}{1-\beta}\left((l\!+\!k_0)^{1-\beta}\!\!-\!\!(k\!+\!k_0\!+\!1)^{1-\beta}\!\right)\! \right)\!\right), &\!\!\!\!\beta\in\!(1/2,1).
		\end{cases}
	\end{align}
\end{lemmax}
\textit{Proof}:
By $\ln(1-x)\leq -x$, $\forall x\in(0,1)$, we have
\begin{align*}
	\prod_{i=l}^k \left(1-\frac{\alpha}{(i+k_0)^\beta}\right)
	= &\exp\left( \sum_{i=l}^k \ln \left( 1-\frac{\alpha}{(i+k_0)^\beta} \right) \right)\\
	\leq &\exp\left( -\sum_{i=l}^k \frac{\alpha}{(i+k_0)^\beta} \right).
\end{align*}
Note that $f(x)=\frac{\alpha}{x+k_0}$ with $\alpha>0$ is a strictly decreasing function for $x>0$. Then, when $\beta=1$, we have
\begin{align*}
	&\exp\left( -\sum_{i=l}^k \frac{\alpha}{i+k_0} \right)
	\leq \exp\left( -\int_l^{k} \frac{\alpha}{x+k_0}\text{d}x \right)\\
	=& \exp\left( \alpha\ln (l+k_0) - \alpha \ln (k+k_0) \right)
	= \left( \frac{l+k_0}{k+k_0} \right)^\alpha.
\end{align*}
When $\beta<1$, from (35) in \cite{Liu2020} it follows that
\begin{align*}
	&\exp\left( -\sum_{i=l}^k \frac{\alpha}{(i+k_0)^\beta} \right)\\
	\leq& \exp\left( \frac{\alpha}{1-\beta}\left((l+k_0)^{1-\beta}-(k+k_0+1)^{1-\beta}\right) \right).
\end{align*}
This completes the proof of \eqref{ineq:prod}.

Note that
\begin{align}\label{eq:prod*prod}
	&\prod_{i=l}^k \left(1-\frac{\alpha}{(i+k_0)^\beta}+\frac{\gamma}{(i+k_0)^{2\beta}}\right)\nonumber\\
	=& \prod_{i=l}^k \left(1-\frac{\alpha}{(i+k_0)^\beta}\right) \prod_{i=l}^k \left(1+O\left(\frac{1}{(i+k_0)^{2\beta}}\right)\right).
\end{align}
Since $\beta>1/2$, by Theorem 2.1.3 of \cite{Pan2015Order}, we have
\begin{align*}
	\sup_{l,k}\prod_{i=l}^k \left(1+O\left(\frac{1}{(i+k_0)^{2\beta}}\right)\right)<\infty,
\end{align*}
which together with \eqref{ineq:prod} and \eqref{eq:prod*prod} implies \eqref{est:prod}.

\begin{lemmax}\label{lemma:calc_sum_exp}
For any given $c, \ k_0\geq 0,\ 0\leq p\leq 1,$ and $q\in \mathbb{R}$, we have
\begin{align*}
		\sum_{l=1}^{k} \frac{\exp\left(c(l+k_0)^p\right)}{(l+k_0)^q} =  O\left( \frac{\exp\left( c(k+k_0)^p \right)}{(k+k_0)^{p+q-1}} \right) .
\end{align*}
\end{lemmax}
\textit{Proof}:
Note that
\begin{align*}
	&\sum_{l=1}^{k}(l+k_0)^{p-1}\exp\left(c(l+k_0)^p\right)\\
	=& O\left( \int_{1+k_0}^{k+k_0} t^{p-1} \exp(ct^p) \text{d}t \right)\\
	=& O\left( \exp\left( c(k+k_0)^p \right) \right).
\end{align*}
Then, using the Abel's transformation (see (6.29) in \cite{Zorich2016}), we have
\begin{align*}
	&\sum_{l=1}^{k} \frac{\exp\left(c(l+k_0)^p\right)}{(l+k_0)^q}\\
	=& \left( \sum_{i=1}^{k} \frac{\exp\left(c(i+k_0)^p\right)}{(i+k_0)^{1-p}} \right) \frac{1}{(k+k_0)^{p+q-1}}\\
	&+ \sum_{l=1}^{k-1} \left( \sum_{i=1}^{l} \frac{\exp\left(c(i+k_0)^p\right)}{(i+k_0)^{1-p}} \right)\\
	&\qquad\cdot\left( \frac{1}{(l+k_0)^{p+q-1}} - \frac{1}{(l+k_0+1)^{p+q-1}} \right)\\
	=& O\left( \frac{\exp\left( c(k+k_0)^p \right)}{(k+k_0)^{p+q-1}} \right) + O\left( \sum_{l=1}^{k} \frac{\exp\left(c(l+k_0)^p\right)}{(l+k_0)^{p+q}} \right),
\end{align*}
which together with
\begin{align*}
	O\left( \sum_{l=1}^{k} \frac{\exp\left(c(l+k_0)^p\right)}{(l+k_0)^{p+q}} \right)
	= o\left( \sum_{l=1}^{k} \frac{\exp\left(c(l+k_0)^p\right)}{(l+k_0)^{q}} \right)
\end{align*}
implies the lemma. $\hfill\Box$

Next, regarding the algorithm's step-size and unlike Assumption \ref{Assumption step}, we further analyze the mean-square and almost-sure convergence rate of the algorithm with the specific step-size form. First, we give the mean-square convergence rate of the algorithm.
\begin{theorem}
Suppose Assumption \ref{Assumption_Graph} holds. Let the step-size $\alpha(k)=\frac{\alpha_k}{(k+a_{2})^\beta}$, $ 0<\underline{\alpha}<\alpha_k<\overline{\alpha}<\infty$, $a_{2}\geq 0$, $\beta\in(0,1]$, $b(k)=O(k^\gamma), \gamma<\beta-\frac{1}{2}$. Then, the mean-square convergence rate of the algorithm is given as follows.\\
When $\beta\in(0,1),$ for all $i, j\in\mathcal{V}$, we have
\begin{eqnarray}\label{them31}
\mathbb{E}\left[x_i(k)-s_{i}x^\star\right]^{2}= O\left( k^{1+2\gamma-2\beta}\right).
\end{eqnarray}
When $\beta=1$, for all $i,j\in\mathcal{V}$, we have
\begin{eqnarray}\label{them32}
&&\mathbb{E}\left[x_i(k)-s_{i}x^\star\right]^{2}\cr
&=&\left\{
\begin{array}{lcc}
O\left(k^{-2\underline{\alpha}\lambda_2(\mathcal{L})}\right),  && \gamma+\underline{\alpha}\lambda_2(\mathcal{L})<1/2;  \\
 O\left(k^{2\gamma-1}\ln k\right),  && \gamma+\underline{\alpha}\lambda_2(\mathcal{L})=1/2;   \\
 O\left(k^{2\gamma-1}\right), && \gamma+\underline{\alpha}\lambda_2(\mathcal{L})>1/2.   \\
  \end{array}
  \right.
\end{eqnarray}
\end{theorem}
\textit{Proof}: For analyzing the mean-square convergence rate of the algorithm, we do it in the following three steps.

{\bf Step 1}: We give the mean-square convergence rate of $s_{i}x_i(k)-\frac{1}{N}\textbf{1}_{N}^Tz(k)$. When $\alpha(k)=\frac{\alpha_k}{(k+a_{2})^\beta}$ and $b(k)=O(k^\gamma), \beta\in(0,1],\gamma<\beta-\frac{1}{2}$, there exists $\varrho>0$ such that
\begin{eqnarray}\label{the35}
2\alpha^{2}(k)b^{2}(k)N\|I_{N}-J\|^{2}\|S\|^{2}\|\mathcal{A}\|^{2}\leq
\frac{\varrho}{(k+a_{2})^{2\beta-2\gamma}}.
\end{eqnarray}
Then, from (\ref{22}) and (\ref{the35}) it follows that
\begin{eqnarray*}
\!\mathbb{E}\big[V(k+1)\big]
&\leq&\!\!\!\left(1-\frac{2\alpha_k\lambda_2(\mathcal{L})}{(k+a_{2})^\beta}+\frac{\alpha_k^{2}\|\mathcal{L}_{S}\|^{2}}{(k+a_{2})^{2\beta}}\right)V(k)\cr
&&\!+\frac{\varrho}{(k+a_{2})^{2\beta-2\gamma}}, ~~{\rm as}~~k>k_{0}.
\end{eqnarray*}
Iterating the above process gives
\begin{eqnarray}\label{30}
&&\!\!\!\!\mathbb{E}\big[V(k+1)\big]\cr
&\leq&\!\!\!\!\!\!\prod\limits_{t=k_{0}}^{k}\left(1-\frac{2\underline{\alpha}\lambda_2(\mathcal{L})}{(t+a_{2})^\beta}+
\frac{\overline{\alpha}^{2}\|\mathcal{L}_{S}\|^{2}}{(t+a_{2})^{2\beta}}\right)\mathbb{E}\left[V(k_{0})\right]\cr
&&\!\!\!\!\!\!+\sum_{l=k_{0}}^{k-1}\!\prod_{t=l+1}^{k}\!\left(1\!-\!\frac{2\underline{\alpha}\lambda_2(\mathcal{L})}{(t+a_{2})^\beta}
\!+\!\frac{\overline{\alpha}^{2}\|\mathcal{L}_{S}\|^{2}}{(t+a_{2})^{2\beta}}\right)
\frac{\varrho}{(l+a_{2})^{2\beta-2\gamma}}\cr
&&\!\!\!\!\!\!+\frac{\varrho}{(k+a_{2})^{2\beta-2\gamma}}.
\end{eqnarray}
When $\beta=1$, from (\ref{est:prod}) and (\ref{30}) it follows that
\begin{eqnarray*}
&&\!\!\!\!\mathbb{E}\big[V(k+1)\big]\cr
&=& O\left(\frac{1}{(k+a_{2})^{2\underline{\alpha}\lambda_2(\mathcal{L})}}\right)+\frac{\varrho}{(k+a_{2})^{2-2\gamma}}
\cr
&&+\sum_{l=k_{0}}^{k-1}\left(\frac{l+a_{2}+1}{(k+a_{2})}\right)^{2\underline{\alpha}\lambda_2(\mathcal{L})}
\frac{\varrho}{(l+a_{2})^{2-2\gamma}}\cr
&=& O\left(\frac{1}{(k+a_{2})^{2\underline{\alpha}\lambda_2(\mathcal{L})}}\right)+O\left(\frac{1}{(k+a_{2})^{2-2\gamma}}
\right)\cr
&&+\frac{\varrho}{(k+a_{2})^{2\underline{\alpha}\lambda_2(\mathcal{L})}}\sum_{l=k_{0}}^{k-1}
\frac{(l+a_{2}+1)^{2\underline{\alpha}\lambda_2(\mathcal{L})}}{(l+a_{2})^{2-2\gamma}}\cr
&=&O\left(\frac{1}{(k+a_{2})^{2\underline{\alpha}\lambda_2(\mathcal{L})}}\right)+O\left(\frac{1}{(k+a_{2})^{2-2\gamma}}
\right)\cr
&&+O\left(\frac{1}{(k+a_{2})^{2\underline{\alpha}\lambda_2(\mathcal{L})}}\sum_{l=k_{0}}^{k-1}
\frac{1}{(l+a_{2})^{2-2\gamma-2\underline{\alpha}\lambda_2(\mathcal{L})}}\right).
\end{eqnarray*}
Note that
\begin{eqnarray*}
&&\sum\limits_{l=k_{0}}^{k-1}\frac{1}{(l+a_{2})^{2-2\gamma-2\underline{\alpha}\lambda_2(\mathcal{L})}}\cr
&\leq&\int_{k_{0}-1}^{k}\frac{1}{(x+a_{2})^{2-2\gamma-2\underline{\alpha}\lambda_2(\mathcal{L})}}\textrm{d}x.
\end{eqnarray*}
Then, we have
\begin{eqnarray*}
&&\sum_{l=k_{0}}^{k-1}\frac{1}{(l+a_{2})^{2-2\gamma-2\underline{\alpha}\lambda_2(\mathcal{L})}}\cr
&=&\left\{
\begin{array}{lcc}
O\left(1\right),  & \gamma+\underline{\alpha}\lambda_2(\mathcal{L})<1/2;  \\
 O\left(\ln k\right),  & \gamma+\underline{\alpha}\lambda_2(\mathcal{L})=1/2;   \\
 O\left((k+a_{2})^{2\gamma+2\underline{\alpha}\lambda_2(\mathcal{L})-1}\right), & \gamma+\underline{\alpha}\lambda_2(\mathcal{L})>1/2.   \\
  \end{array}
  \right.
\end{eqnarray*}
Thus, when $\beta=1$, for all $i,j\in\mathcal{V}$, we have
\begin{eqnarray}\label{them33}
&&\!\!\!\!\!\!\mathbb{E}\big[V(k+1)\big]\cr
&=&\!\!\!\!\!\!\left\{
\begin{array}{lcl}
O\left((k+a_{2})^{-2\underline{\alpha}\lambda_2(\mathcal{L})}\right),  & \gamma+\underline{\alpha}\lambda_2(\mathcal{L})<1/2;  \\
 O\left((k+a_{2})^{2\gamma-1}\ln k\right),  & \gamma+\underline{\alpha}\lambda_2(\mathcal{L})=1/2;   \\
 O\left((k+a_{2})^{2\gamma-1}\right), & \gamma+\underline{\alpha}\lambda_2(\mathcal{L})>1/2.   \\
  \end{array}
  \right.
\end{eqnarray}
When $0<\beta<1$, there exists $k>k_{0}$ such that
\begin{eqnarray*}
-\frac{2\underline{\alpha}\lambda_2(\mathcal{L})}{(k+a_{2})^\beta}+\frac{\overline{\alpha}^{2}\|\mathcal{L}_{S}\|^{2}}{(k+a_{2})^{2\beta}}\leq -\frac{\underline{\alpha}\lambda_2(\mathcal{L})}{(k+a_{2})^\beta},
\end{eqnarray*}
and from (\ref{30}) it follows that
\begin{eqnarray*}
&&\!\!\!\!\mathbb{E}\big[V(k+1)\big]\cr
&\leq&\prod\limits_{t=k_{0}}^{k}\left(1-\frac{\underline{\alpha}\lambda_2(\mathcal{L})}{(t+a_{2})^\beta}\right)\mathbb{E}\left[V(k_{0})\right]
+\frac{\varrho}{(k+a_{2})^{2\beta-2\gamma}}\cr
&&\!\!\!\!+\sum_{l=k_{0}}^{k-1}\prod_{t=l+1}^{k}\left(1-\frac{\underline{\alpha}\lambda_2(\mathcal{L})}{(t+a_{2})^\beta}\right)
\frac{\varrho}{(l+a_{2})^{2\beta-2\gamma}}.
\end{eqnarray*}
Note that $(1-\frac{\underline{\alpha}\lambda_2(\mathcal{L})}{(l+a_{2})^\beta})^{-1}\leq 2$ holds if $l\geq k_0$ for sufficiently large $k_0$. Then, from (\ref{ineq:prod}) it follows that
\begin{eqnarray}\label{34}
&&\!\!\!\!\mathbb{E}\big[V(k+1)\big]\cr
&\leq&\prod\limits_{t=k_{0}}^{k}\left(1-\frac{\underline{\alpha}\lambda_2(\mathcal{L})}{(t+a_{2})^\beta}\right)\mathbb{E}\big[V(k_{0})\big]
+\frac{\varrho}{(k+a_{2})^{2\beta-2\gamma}}\cr
&&\!\!\!\!+2\sum_{l=k_{0}}^{k-1}\prod_{t=l}^{k}\left(1-\frac{\underline{\alpha}\lambda_2(\mathcal{L})}{(t+a_{2})^\beta}\right)
\frac{\varrho}{(l+a_{2})^{2\beta-2\gamma}}\cr
&=&\!\!O\left(\exp\left(-\frac{\underline{\alpha}\lambda_2(\mathcal{L})}{1-\beta}(k+a_{2}+1)^{1-\beta}\right)\right)\cr
&&\!\!+O\left(\frac{1}{(k+a_{2})^{2\beta-2\gamma}}\right)\cr
&&\!\!+O\left(\sum_{l=k_{0}}^{k-1}\exp\left(-\frac{\underline{\alpha}\lambda_2(\mathcal{L})}{1-\beta}(k+a_{2}+1)^{1-\beta}\right)\right.\cr
&&\!\!\left.\times\exp\left(\frac{\underline{\alpha}\lambda_2(\mathcal{L})}{1-\beta}(l+a_{2})^{1-\beta}\right)\frac{\varrho}{(l+a_{2})^{2\beta-2\gamma}}\right).
\end{eqnarray}
Note that for large $k_0$, $\frac{\beta-2\gamma}{\underline{\alpha}\lambda_2(\mathcal{L})(k_{0}+a_{2})^{1-\beta}}<\frac{1}{2}$. Then, we have
\begin{eqnarray*}
&&\!\!\!\!\!\!\sum_{l=k_{0}}^{k-1}\exp\left(\frac{\underline{\alpha}\lambda_2(\mathcal{L})}{1-\beta}(l+a_{2})^{1-\beta}\right)\frac{\varrho}{(l+a_{2})^{2\beta-2\gamma}}\cr
&\!\!\!\!\!\!\leq&\!\!\!\!\int_{k_{0}}^{k}\exp\left(\frac{\underline{\alpha}\lambda_2(\mathcal{L})}{1-\beta}(l+a_{2})^{1-\beta}\right)\frac{\varrho}{(l+a_{2})^{2\beta-2\gamma}}\textrm{d}l\cr
&\!\!\!\!\!\!=&\!\!\!\!\!\!\frac{1}{\underline{\alpha}\lambda_2(\mathcal{L})}\int_{k_{0}}^{k}\frac{\varrho}{(l+a_{2})^{\beta-2\gamma}}\textrm{d}\left(\exp\left(\frac{\alpha_t\lambda_2(\mathcal{L})}{1-\beta}(l+a_{2})^{1-\beta}\right)\right)\cr
&\!\!\!\!\!\!=&\!\!\!\!\!\!\left.\frac{1}{\underline{\alpha}\lambda_2(\mathcal{L})}\frac{\varrho}{(l+a_{2})^{\beta-2\gamma}}
\exp\left(\frac{\underline{\alpha}\lambda_2(\mathcal{L})}{1-\beta}(l+a_{2})^{1-\beta}\right)\right|^{k}_{k_{0}}\cr
&\!\!\!\!\!\!&\!\!\!\!\!\!-\frac{\varrho}{\underline{\alpha}\lambda_2(\mathcal{L})}\int_{k_{0}}^{k}\exp\left(\frac{\underline{\alpha}\lambda_2(\mathcal{L})}{1-\beta}(l+a_{2})^{1-\beta}\right)
\textrm{d}\left(\frac{1}{(l+a_{2})^{\beta-2\gamma}}\right)\cr
&\!\!\!\!\!\!\leq&\!\!\!\!\!\!\frac{1}{\underline{\alpha}\lambda_2(\mathcal{L})}\frac{\varrho}{(k+a_{2})^{\beta-2\gamma}}
\exp\left(\frac{\underline{\alpha}\lambda_2(\mathcal{L})}{1-\beta}(k+a_{2})^{1-\beta}\right)\cr
&\!\!\!\!\!\!&\!\!\!\!+\frac{\beta-2\gamma}{\underline{\alpha}\lambda_2(\mathcal{L})(k_{0}+a_{2})^{1-\beta}}\int_{k_{0}}^{k}
\exp\left(\frac{\underline{\alpha}\lambda_2(\mathcal{L})}{1-\beta}(l+a_{2})^{1-\beta}\right)\cr
&&\qquad\qquad\qquad\qquad\qquad\qquad\qquad\cdot\frac{\varrho}{(l+a_{2})^{2\beta-2\gamma}}\textrm{d}l\cr
&\!\!\!\!\!\!\leq&\!\!\!\!\!\!\frac{1}{\underline{\alpha}\lambda_2(\mathcal{L})}\frac{\varrho}{(k+a_{2})^{\beta-2\gamma}}
\exp\left(\frac{\underline{\alpha}\lambda_2(\mathcal{L})}{1-\beta}(k+a_{2})^{1-\beta}\right)\cr
&\!\!\!\!\!\!&\!\!\!\!+\frac{1}{2}\int_{k_{0}}^{k}
\exp\left(\frac{\underline{\alpha}\lambda_2(\mathcal{L})}{1-\beta}(l+a_{2})^{1-\beta}\right)
\frac{\varrho}{(l+a_{2})^{2\beta-2\gamma}}\textrm{d}l.
\end{eqnarray*}
Furthermore, we have
\begin{eqnarray*}
&&\sum_{l=k_{0}}^{k-1}\exp\left(\frac{a_{1}\lambda_2(\mathcal{L})}{1-\beta}(l+a_{2})^{1-\beta}\right)\frac{\varrho}{(l+a_{2})^{2\beta-2\gamma}}\cr
&=&O\left(\frac{1}{(k+a_{2})^{\beta-2\gamma}}
\exp\left(\frac{\underline{\alpha}\lambda_2(\mathcal{L})}{1-\beta}(k+a_{2})^{1-\beta}\right)\right).
\end{eqnarray*}
From (\ref{34}) it follows that
\begin{eqnarray}\label{them34}
&&\!\!\!\!\mathbb{E}\big[V(k+1)\big]\cr
&=&\!\!O\left(\exp\left(-\frac{\underline{\alpha}\lambda_2(\mathcal{L})}{1-\beta}(k+a_{2}+1)^{1-\beta}\right)\right)\cr
&&\!\!+O\left(\frac{1}{(k+a_{2})^{2\beta-2\gamma}} \right)\cr
&&\!\!+O\left(\exp\left(-\frac{\underline{\alpha}\lambda_2(\mathcal{L})}{1-\beta}(k+a_{2}+1)^{1-\beta}\right)\right.\cr
&&\!\!\left.\cdot\frac{1}{(k+a_{2})^{\beta-2\gamma}}
\exp\left(\frac{\underline{\alpha}\lambda_2(\mathcal{L})}{1-\beta}(k+a_{2})^{1-\beta}\right)\right).\cr
&=& O\left((k+a_{2})^{2\gamma-\beta}\right).
\end{eqnarray}

{\bf Step 2}: We give the mean-square convergence rate of  $\frac{1}{N}\textbf{1}_{N}^Tz(k)-x^\star$. From (\ref{eq:sum ave z}) it follows that
\begin{eqnarray*}
&&\mathbb{E}\left[\frac{1}{N}\textbf{1}_{N}^Tz(k)-x^\star\right]^2\cr
&=&\mathbb{E}\left[-\frac{1}{N}\sum_{j=k+1}^{\infty}\alpha(j-1) ({{\bf{1}}_{N}^T} S\mathcal{A})\omega(j-1)\right]^2\cr
&=&\frac{2\sum_{i\in \mathcal{V}} c_i^2}{N^{2}}\sum_{j=k+1}^\infty \alpha^2(j)b^2(j)\cr
&=&O\left(\sum_{j=k+1}^\infty \frac{1}{(j+a_{2})^{2\beta-2\gamma}}\right).
\end{eqnarray*}
Note that $\gamma<\beta-\frac{1}{2}$ and
\begin{eqnarray*}
\sum_{j=k+1}^\infty \frac{1}{(j+a_{2})^{2\beta-2\gamma}}
\leq\int_{k}^{\infty}\frac{1}{(x+a_{2})^{2\beta-2\gamma}}\textrm{d}x.
\end{eqnarray*}
Then, we have
\begin{eqnarray*}
\sum_{j=k+1}^\infty\frac{1}{(j+a_{2})^{2\beta-2\gamma}}=O\left((k+a_{2})^{1+2\gamma-2\beta}\right),
\end{eqnarray*}
which implies that
\begin{eqnarray}\label{them35}
\mathbb{E}\left[\frac{1}{N}\textbf{1}_{N}^Tz(k)-x^\star\right]^2=O\left((k+a_{2})^{1+2\gamma-2\beta}\right).
\end{eqnarray}

{\bf Step 3}: We give the mean-square convergence rate of the algorithm. Note that
\begin{eqnarray}\label{them36}
\!\!\!\!&&\!\!\!\!\mathbb{E}\left[x_i(k)-s_{i}x^\star\right]^{2}\cr
\!\!\!\!&=&\!\!\!\!\mathbb{E}\left[s_{i}x_i(k)-\frac{1}{N}\textbf{1}_{N}^Tz(k)+\frac{1}{N}\textbf{1}_{N}^Tz(k)-x^\star\right]^2\cr
\!\!\!\!&\leq&\!\!\!\!2\mathbb{E}\left[s_{i}x_i(k)-\frac{1}{N}\textbf{1}_{N}^Tz(k)\right]^{2}\!\!+
2\mathbb{E}\left[\frac{1}{N}\textbf{1}_{N}^Tz(k)-x^\star\right]^2.
\end{eqnarray}
Then, when $\beta=1$, from (\ref{them33}), (\ref{them35}) and (\ref{them36}) it follows that (\ref{them32}) holds; when $0<\beta<1$, from (\ref{them34}), (\ref{them35}) and (\ref{them36}) it follows that (\ref{them31}) holds. The proof is completed. $\hfill\Box$

In the following, we give the almost-sure convergence rate of the algorithm with the specific step-size form.
\begin{theorem}
Suppose Assumption \ref{Assumption_Graph} holds. Let the step-size $\alpha(k)=\frac{\alpha_k}{(k+a_{2})^\beta}$, $0<\underline{\alpha}<\alpha_k<\overline{\alpha}<\infty$, $a_{2}\geq 0$, $\beta\in(1/2,1]$, $b(k)=O(k^\gamma), \gamma<\beta-1/2$. Then, the almost-sure convergence rate of the algorithm is given as follows.\\
When $\beta\in(1/2,1),$ for any $\eta\in\left(\frac{1}{4},\frac{\beta/2-\gamma}{1-\beta}\right)$ and all $i, j\in\mathcal{V}$, we have
\begin{eqnarray}\label{them41}
s_i x_i(k)-s_j x_j(k) = O\left( k^{\gamma+\eta-(\eta+1/2)\beta} \right),\quad \text{a.s.}
\end{eqnarray}
When $\beta=1$, for all $i,j\in\mathcal{V}$, we have
\begin{eqnarray}\label{them42}
&&\!\!\!\!\!\!s_i x_i(k) - s_j x_j(k)\nonumber\\
&\!\!\!\!\!\!\!\!\!\!=&\!\!\!\!\!\!\!\!\! \begin{cases}
				O\!\!\left( k^{\gamma-1/2}\sqrt{\ln \ln k} \right), &\!\!\!\! \underline{\alpha}\lambda_2(\mathcal{L})\!+\!\gamma\!>\!1/2;\\
				O\!\!\left(k^{\gamma-1/2} \sqrt{\ln k \ln \ln \ln k} \right), &\!\!\!\! \underline{\alpha}\lambda_2(\mathcal{L})\!+\!\gamma\!=\!1/2;\\
				O\!\!\left( k^{-\underline{\alpha}\lambda_2(\mathcal{L})} \right), &\!\!\!\! \underline{\alpha}\lambda_2(\mathcal{L})\!+\!\gamma\!<\!1/2.
			\end{cases}  \text{a.s.}
\end{eqnarray}
\end{theorem}
\textit{Proof}:	
As clarified in Theorem \ref{Theorem_Convergence}, it is equivalent to calculate the convergence rate of $\delta(k)= z(k)-Jz(k)$, where $z(k)$ and $J$ are defined in Theorem \ref{Theorem_Convergence}.

Note that
\begin{align*}
	\lVert\delta(k)\rVert
	=&\sup_{\lVert v\rVert=1} \lvert v^T \delta(k)\rvert
	=\sup_{\substack{p^2\lVert e\rVert^2 + N q^2 = 1 \\ e^T\mathbf{1}=0}} \lvert (p e + q \mathbf{1})v^T \delta(k)\rvert\\
	=&\sup_{\substack{p^2\lVert e\rVert^2 + Nq^2 = 1 \\ e^T\mathbf{1}=0}} \lvert p e^T z(k) \rvert
	=\sup_{\substack{\lVert e\rVert=1 \\ e^T\mathbf{1}=0}} \lvert e^T z(k) \rvert,
\end{align*}
and there exists $e_1, \ldots, e_{N-1} $ such that $e_i^T e_j =0 $ for $i\neq j $, $\lVert e_i\rVert =1 $ and $e_i^T \mathbf{1}=0 $ for $i=1,2,\ldots,N-1$. Then, we have
\begin{align*}
	&\sup_{\substack{\lVert e\rVert=1 \\ e^T\mathbf{1}=0}} \lvert e^T z(k) \rvert
	=\sup_{\sum_{i=l}^{N-1}p_{i}^2=1} \sum_{i=l}^{N-1} \lvert p_i\rvert \lvert e_i^T z(k) \rvert\\
	\leq& \sup_{\sum_{i=l}^{N-1}p_{i}^2=1} \sqrt{N\sum_{i=l}^{N-1} \lvert p_i\rvert^2} \max_{i} \lvert e_i^T z(k) \rvert\\
	=& \sqrt{N}\max_{i} \lvert e_i^T z(k)\rvert.
\end{align*}
Set
\begin{align*}
	\tilde{\mathcal{D}}=\begin{bmatrix}
		e_1 & \cdots & e_{N-1}
	\end{bmatrix}^T, \quad
	\mathcal{D} =\begin{bmatrix}
		e_1 & \cdots & e_{N-1} & \frac{\mathbf{1}}{\sqrt{N}}
	\end{bmatrix}^T.
\end{align*}
Then, to calculate the convergence rate of $\delta(k)$, it suffices to analyze that of $\tilde{\mathcal{D}} z(k)$.

From the properties of $ e_i $, we have $ \mathcal{D}^T \mathcal{D} = I_N $. Let  $\tilde{\mathcal{L}}=\tilde{\mathcal{D}} \mathcal{L}_{S} \tilde{\mathcal{D}}^T $. Then, $\mathcal{D} \mathcal{L}_{S} \mathcal{D}^T = \rm{diag}(\tilde{\mathcal{L}},0)$ and $\lambda_{\min}(\tilde{\mathcal{L}}) = \lambda_2(\mathcal{L}).$

From $\alpha(k)=\frac{\alpha_k}{(k+a_{2})^\beta}$ and (\ref{consensusstate}) it follows that
\begin{align*}
z(k+1)
\!\!=\!\!&\left(I-\frac{\alpha_k}{(k+a_{2})^\beta}\mathcal{L}_{S}\right)z(k)\!\!+\!\!\frac{\alpha_k b(k)}{(k+a_{2})^\beta}S\mathcal{A}\frac{\omega(k)}{b(k)}.
\end{align*}
Hence, we have
\begin{align*}
&\mathcal{D}z(k+1)\\
=\!\!&\left( I_N\!\!-\!\! \frac{\alpha_k}{(k+a_{2})^\beta}\mathcal{D}\mathcal{L}_{S}\mathcal{D}^T\right)\mathcal{D}z(k)\!\!+\!\!\frac{\alpha_k b(k)}{(k+a_{2})^\beta}\mathcal{D}S\mathcal{A}\frac{\omega(k)}{b(k)}\\
=\!\!&\begin{bmatrix}
		I_{N-1}\!\!-\!\! \frac{\alpha_k}{(k+a_{2})^\beta}\tilde{\mathcal{L}} & 0 \\ 0 & 1
	\end{bmatrix}
	\mathcal{D} z(k)\!\!+\!\!\frac{\alpha_k b(k)}{(k+a_{2})^\beta} \mathcal{D} S\mathcal{A}\frac{\omega(k)}{b(k)},
\end{align*}
which implies
\begin{align*}
	&\tilde{\mathcal{D}} z(k+1)\\
	=&\!\!\left( I_{N-1}\!\!-\!\!\frac{\alpha_k}{(k+a_{2})^\beta}\tilde{\mathcal{L}} \right)\tilde{\mathcal{D}} z(k)\!\!+\!\!\frac{\alpha_k b(k)}{(k+a_{2})^\beta} \tilde{\mathcal{D}} S\mathcal{A}\frac{\omega(k)}{b(k)}
\end{align*}
Iterating the above equation results in
\begin{align}\label{eq:tilde(D) z}
&\tilde{\mathcal{D}}z(k) \nonumber\\
\!\!=\!\!&\sum_{l=0}^{k-1} \prod_{i=l+1}^{k-1} \left(I\!\!-\!\! \frac{\alpha_i}{(i+a_{2})^\beta}\tilde{\mathcal{L}}\right) \frac{\alpha_l b(l)}{(l+a_{2})^\beta} \tilde{\mathcal{D}}S\mathcal{A}\frac{w(l)}{b(l)} \nonumber\\
	&\!\!+ \!\!\prod_{i=0}^{k-1}\left(I\!\!-\!\! \frac{\alpha_i}{(i+a_{2})^\beta}\tilde{\mathcal{L}}\right)\tilde{\mathcal{D}}z(0).
\end{align}

Note that when $\beta<1$, by Lemma \ref{lemma:prod_est}, we have
\begin{align}\label{eq:prod (I-poly L)}
	&\prod_{i=l+1}^{k-1} \left(I- \frac{\alpha_i}{(i+a_{2})^\beta}\tilde{\mathcal{L}}\right)
	\leq \prod_{i=l+1}^{k-1} \left(1 - \frac{\underline{\alpha}\lambda_{2}(\mathcal{L})}{(i+a_{2})^\beta} \right)\nonumber\\
	= &O\left( \exp\left( \frac{\underline{\alpha}\lambda_2(\mathcal{L})}{1-\beta}\left((l+a_{2})^{1-\beta}-(k+a_{2})^{1-\beta}\right) \right) \right).
\end{align}
According to the Lemma 2 in \cite{Wei1985} and Lemma \ref{lemma:calc_sum_exp}, for any $\eta>\frac{1}{4}$, we have
\begin{align}\label{eq:sum_exp poly}
	&\sum_{l=0}^{k-1}\exp\left( \frac{\underline{\alpha}\lambda_2(\mathcal{L})}{1-\beta}(l+a_{2})^{1-\beta} \right)\frac{\alpha_l b(l)}{(l+a_{2})^\beta} \tilde{\mathcal{D}}S\mathcal{A}\frac{w(l)}{b(l)}\nonumber\\
	=& O\left(\sum_{l=0}^{k-1}\exp\left( \frac{\underline{\alpha}\lambda_2(\mathcal{L})}{1-\beta}(l+a_{2})^{1-\beta} \right)\frac{\alpha_l\tilde{\mathcal{D}}S\mathcal{A}}{(l+a_{2})^{\beta-\gamma}} \frac{w(l)}{b(l)}\right)\nonumber\\
	=& O\left(\exp\left( \frac{\underline{\alpha}\lambda_2(\mathcal{L})}{1-\beta}(k+a_{2})^{1-\beta} \right) (k+a_{2})^{\gamma + \eta-(\eta+1/2)\beta}\right),\nonumber\\
	&\qquad\qquad\qquad\qquad\qquad\qquad\qquad\qquad\qquad \text{a.s.}
\end{align}
Substituting \eqref{eq:prod (I-poly L)} and \eqref{eq:sum_exp poly} into \eqref{eq:tilde(D) z} gives
\begin{align*}
	\tilde{\mathcal{D}}z(k) = O\left( k^{\gamma + \eta-(\eta+1/2)\beta} \right), \ \text{a.s.}
\end{align*}
When $\beta=1$, by Lemma \ref{lemma:prod_est}, we have
\begin{align}\label{eq:prod (I-harm L)}
\prod_{i=l+1}^{k-1} \left(I - \frac{\alpha_i}{i+a_{2}}\tilde{\mathcal{L}}\right)
	\leq& \prod_{i=l+1}^{k-1} \left(1 - \frac{\underline{\alpha}\lambda_2(\mathcal{L})}{i+a_{2}} \right)\nonumber\\
	=& O\left( \left( \frac{l+a_{2}}{k+a_{2}} \right)^{\underline{\alpha}\lambda_2(\mathcal{L})} \right).
\end{align}
According to the Lemma 2 in \cite{Wei1985}, one can get
\begin{align}\label{eq:sum_poly} &\frac{1}{(k+a_{2})^{\underline{\alpha}\lambda_2(\mathcal{L})}}\sum_{l=0}^{k-1}\frac{\alpha_l b(l)}{(l+a_{2})^{1-\underline{\alpha}\lambda_2(\mathcal{L})}} \tilde{\mathcal{D}}S\mathcal{A}\frac{w(l)}{b(l)}\nonumber\\
	=& O\left(\frac{1}{(k+a_{2})^{\underline{\alpha}\lambda_2(\mathcal{L})}}\sum_{l=0}^{k-1}
\frac{\alpha_l\tilde{\mathcal{D}}S\mathcal{A}}{(l+a_{2})^{1-\gamma-\underline{\alpha}\lambda_2(\mathcal{L})}} \frac{w(l)}{b(l)}\right)\nonumber\\
	=& \begin{cases}
		O\left( k^{\gamma-1/2}\sqrt{\ln \ln k} \right), & \underline{\alpha}\lambda_2(\mathcal{L}) +\gamma > 1/2;\\
		O\left(k^{\gamma-1/2} \sqrt{\ln k \ln \ln \ln k} \right), & \underline{\alpha}\lambda_2(\mathcal{L}) +\gamma = 1/2;\\
		O\left( k^{-\underline{\alpha}\lambda_2(\mathcal{L})} \right), & \underline{\alpha}\lambda_2(\mathcal{L}) +\gamma < 1/2,
	\end{cases}\text{a.s.}
\end{align}
Substituting \eqref{eq:prod (I-harm L)} and \eqref{eq:sum_poly} into \eqref{eq:tilde(D) z} gives
\begin{align*}
&\!\! \tilde{\mathcal{D}}z(k)\cr
\!\!=&\!\!\begin{cases}
		O\left( k^{\gamma-1/2}\sqrt{\ln \ln k} \right), & \underline{\alpha}\lambda_2(\mathcal{L}) +\gamma > 1/2;\\
		O\left(k^{\gamma-1/2} \sqrt{\ln k \ln \ln \ln k} \right), & \underline{\alpha}\lambda_2(\mathcal{L}) +\gamma = 1/2;\\
		O\left( k^{-\underline{\alpha}\lambda_2(\mathcal{L})} \right), & \underline{\alpha}\lambda_2(\mathcal{L}) +\gamma < 1/2,
	\end{cases}\text{a.s.}
\end{align*}
This completes the proof.  $\hfill\Box$
\begin{remark}
Communication imperfections in networked systems can be modeled as communication noises  \cite{Li2009,Li2010CommunicationNoise,Huang2010,Huang2012,Huang2015}, which can be regarded as the differential privacy-noise considered here. Therefore, Algorithm 1 can also be used to counteract such communication imperfections in distributed consensus.
\end{remark}
\subsection{Privacy analysis}
This subsection demonstrates that the algorithm is $\epsilon$-differentially private on dataset $D=\{x_i(0), i\in \mathcal{V}\}$. Before giving the privacy analysis, we first introduce the definition of sensitivity. For a private dataset $D$ and an observation $O=\{y_i(k-1), i\in \mathcal{V}\}_{k=1}^T$, there exists a determinate sequence of noises $\{\omega_i(k-1), i\in \mathcal{V}\}_{k=1}^T$ and a determinate trajectory $\rho(D,O)=\{x_i^{D,O}(k-1), i\in \mathcal{V}\}_{k=1}^T$. Below we first give the sensitivity of the algorithm.
\begin{definition}(Sensitivity).
The sensitivity with respect to a randomized mechanism $\mathcal{M}$ at time $k\ge 1$ is given as follows.
\begin{align*}
\scalebox{0.95}{$S(k) =\sup\limits_{D,D'\in \mathcal{D}, O \in\mathcal{O}}\|\rho(D,O)(k-1)-\rho(D',O)(k-1)\|_{1}.$}
\end{align*}
\end{definition}

Sensitivity is a measure of the difference of two trajectories induced by changing the private dataset.
\begin{theorem}  \label{Sensitivity}
Suppose Assumption \ref{Assumption_Graph} holds. Then, the sensitivity of the algorithm is
\begin{align} \label{TSensitivity}
S(k) \le \left\{ \begin{array}{l}
\delta, ~k=1;\\
 \prod_{l=0}^{k-2} (1-\alpha(l)c_{\min})\delta, ~k\ge 2.
\end{array} \right.
\end{align}
\end{theorem}
\textit{Proof}:
Denote $\mathcal{P}=\{\rho(D,O): O\in\mathcal{O}\}$ and $\mathcal{P}'=\{\rho(D',O): O\in\mathcal{O}\}$ as the sets of possible trajectories under the controller (\ref{controller}) w.r.t. $D$ and $D'$ in the observation set $\mathcal{O}$, and the trajectories subject to the probability density functions $f(D,\rho(D,O))$ and $f(D',\rho(D',O))$, respectively.
Based on the controller (\ref{controller}), we have
\begin{align*}
x^{D,O}_i(k)
=& (1-\alpha(k-1)c_i)x^{D,O}_i(k-1) \nonumber\\
&+\alpha(k-1)\sum_{j\in \mathcal{N}_i}|a_{ij}|\text{sgn}(a_{ij})y_j(k-1).
\end{align*}
Similarly, for $D'$ we have
\begin{align*}
x^{D',O}_i(k)
=& (1-\alpha(k-1)c_i)x^{D',O}_i(k-1)  \nonumber\\
&+\alpha(k-1)\sum_{j\in \mathcal{N}_i} |a_{ij}|\text{sgn}(a_{ij})y_j(k-1).
\end{align*}
Since observations $O=\{y_i(k-1), i\in \mathcal{V}\}$ for $D$ and $D'$ are the same, we have
\begin{align*}
& x^{D',O}_i(k)-x^{D,O}_i(k)\nonumber\\
=& (1-\alpha(k-1)c_i)\left(x^{D',O}_i(k-1)-x^{D,O}_i(k-1)\right) \nonumber\\
=& \prod_{l=0}^{k-1} (1-\alpha(l)c_i)\left(x^{D',O}_i(0)-x^{D,O}_i(0)\right).
\end{align*}
Thus, it follows that for $k=1$
\begin{align} \label{Formula_XIsolution41}
&\| \rho(D,O)(k-1)-\rho(D',O)(k-1)\|_{1}  \nonumber\\
=~ & \sum_{i\in\mathcal{V}} \left|x^{D',O}_{i} (0) - x^{D,O}_{i}(0) \right|
\le~\delta,
\end{align}
which implies that $S(1)\leq\delta$, and for $k\ge2$
\begin{align} \label{Formula_XIsolution4}
&\|\rho(D,O)(k-1)-\rho(D',O)(k-1)\|_{1} \nonumber\\
=& \sum_{i\in\mathcal{V}} \left|x^{D',O}_{i} (k-1) - x^{D,O}_{i}(k-1) \right| \nonumber\\
=& \sum_{i\in\mathcal{V}} \left( \prod_{l=0}^{k-2} (1-\alpha_{l}c_i)\right) \left|x^{D',O}_{i}(0)-x^{D,O}_{i}(0)\right| \nonumber\\
\le & \left(\prod_{l=0}^{k-2} (1-\alpha_{l}c_{\min}) \right)\sum_{i\in\mathcal{V}} \left|x^{D',O}_{i}(0)-x^{D,O}_{i}(0)\right| \nonumber\\
\le & \left(\prod_{l=0}^{k-2} (1-\alpha_{l}c_{\min}) \right)\delta.
\end{align}
Thus, $S(k)\leq\prod_{l=0}^{k-2}(1-\alpha_{l}c_{\min})\delta$ for $k\ge 2$. This completes the proof.  $\hfill\Box$

Next, we calculate the algorithm's differential privacy level $\epsilon$.

\begin{theorem}\label{EpsilonSensitivity}
Suppose Assumption \ref{Assumption_Graph} holds. Then, the algorithm is $\epsilon$-differentially private over time horizon $T$ with
\begin{align}\label{Tepsilon}
\epsilon=\sum_{k=1}^T\frac{S(k)}{b(k)}.
\end{align}
\end{theorem}
\textit{Proof}:
Recall that $\mathcal{P}\!=\!\{\rho(D,O): O\in\mathcal{O}\}$ and $\mathcal{P}'\!=\!\{\rho(D',O): O\in\mathcal{O}\}$ are the sets of possible trajectories under controller (\ref{controller}) w.r.t. $D$ and $D'$ in the observation set $\mathcal{O}$, and the trajectories subject to the probability density functions $f(D,\rho(D,O))$ and $f(D',\rho(D',O))$, respectively. Then, it can be obtained that
\begin{align*}
\frac{\mathbb{P}[\mathcal{M}(D)\in \mathcal{O}]}{\mathbb{P}[\mathcal{M}(D')\in \mathcal{O}]} = \frac{\int_{\rho(D,O)\in \mathcal{P}} f\left(D,\rho(D,O)\right) \rm{d} \tau }{\int_{\rho(D',O)\in \mathcal{P}'} f\left(D',\rho(D',O)\right) \rm{d} \tau'}
\end{align*}
Let $\mathcal{T}=\{1,2,\ldots,T\}$ and $\mathcal{W}=\mathcal{V}\times\mathcal{T}$. Then, the probability density functions $f(D,\rho(D,O))$ over time horizon $T$ can be expressed as
\begin{align}\label{Formula_PA3}
&f(D,\rho(D,O)) \nonumber\\
=& \prod_{i\in \mathcal{V},~k\in \mathcal{T}} f(D,\rho(D,O)_i(k-1)) \nonumber\\
=&\!\!\! \prod_{(i,k)\in \mathcal{W}} \!\! \frac{1}{2b(k)} \exp\left(-\frac{\left|\rho(D,O)_{i}(k-1)\!\!-\!\!y_{i}(k-1)\right|}{b(k)}\right)
\end{align}
As they have the same observation over time horizon $T$, there exists a bijection $g(\cdot)$$:\!\mathcal{P} \to \mathcal{P}'$, such that for any pair of $\rho(D,O) \in \mathcal{P}$ and $\rho(D',O) \in \mathcal{P}'$, it has $g(\rho(D,O))=\rho(D',O)$. From the rationale of $y_i(k-1)=x_i(k-1)+\omega_{i}(k-1)$, $\omega_i(k-1)\sim \text{Lap}(0, b(k-1))$, and the observations $O=\{y_i(k-1), i\in\mathcal{V}\}_{k=1}^{T}$, by (\ref{Formula_PA3}) we have
\begin{align*}
&\frac{\mathbb{P}[\mathcal{M}(D)\in \mathcal{O}]}{\mathbb{P}[\mathcal{M}(D')\in \mathcal{O}]} = \frac{\int_{\rho(D,O)\in \mathcal{P}} f\left(D,\rho(D,O)\right) \rm{d} \tau }{\int_{g(\rho(D,O))\in \mathcal{P}'} f\left(D',g(\rho(D,O))\right) \rm{d} \tau} \nonumber\\
&= \frac{\int_{\rho(D,O)\in \mathcal{P}} f\left(D,\rho(D,O)\right) \rm{d} \tau }{\int_{\rho(D,O)\in \mathcal{P}} f\left(D',g(\rho(D,O))\right) \rm{d} \tau}\nonumber\\
&=  \prod_{(i,k)\in \mathcal{W}}   \exp\left(-\frac{\left|\rho(D,O)_{i}(k-1)-y_{i}(k-1)\right|}{b(k)}\right.\nonumber\\
&~~~+\left.\frac{\left|\rho(D',O)_{i}(k-1)-y_{i}(k-1)\right|}{b(k)} \right) \nonumber\\
&\le  \prod_{(i,k)\in \mathcal{W}}  \exp\left(\frac{\left|x^{D',O}_{i}(k-1)-x^{D,O}_{i}(k-1)\right| }{b(k)} \right),
\end{align*}
which together with (\ref{Formula_XIsolution4}) leads to
\begin{align*}
&\frac{\mathbb{P}[\mathcal{M}(D)\in \mathcal{O}]}{\mathbb{P}[\mathcal{M}(D')\in \mathcal{O}]}  \nonumber\\
=& \exp\left(\sum_{k\in \mathcal{T}}\frac{ \sum_{i\in \mathcal{V}} \left|x^{D',O}_{i}(k-1)-x^{D,O}_{i}(k-1)\right| }{b(k)} \right) \nonumber\\
\le& \exp\left(\sum_{k=1}^T\frac{S(k)}{b(k)} \right).
\end{align*}
Hence, we can obtain that $\epsilon=\sum_{k=1}^T\frac{ S(k)}{b(k)}$.  $\hfill\Box$

\begin{remark}
Theorem \ref{EpsilonSensitivity} reveals that the differential privacy level $\epsilon$ can be described by the sum of sensitivity. According to (\ref{TSensitivity}), greater $\alpha(k)$ gives a smaller $S(k)$, which further leads to a smaller $\epsilon$, and thus, a better privacy-preserving. Similarly, if we select privacy noises with a greater parameter $b(k)$, then $\epsilon$ becomes smaller, and thus, the privacy-preserving is stronger.
\end{remark}

Next, we focus on how to design the time-varying step-size $\alpha(k)$ and the noise parameter $b(k)$ to satisfy the predefined $\epsilon^\star$-differential privacy over the infinite time horizon. For the convenience of analysis, we first introduce two lemmas.
\begin{lemmax}\label{lemma:Gamma}
For $\gamma<1,\ 0<\beta<1,\ \nu>0$, we have
\begin{align*}
	\int_{1}^{\infty} x^{-\gamma}\exp\left( -\nu x^{1-\beta} \right){\rm{d}}x
	= \frac{\nu^{-\frac{1-\gamma}{1-\beta}}}{1-\beta} \Gamma\left(\frac{1-\gamma}{1-\beta},\nu\right),
\end{align*}
where $\Gamma(\cdot,\cdot)$ is the upper incomplete gamma function.
\end{lemmax}

\textit{Proof}:
Denote $t=\nu x^{1-\beta}$. Then, we have $\textrm{d}t=\nu(1-\beta)x^{-\beta}\textrm{d}x$, and
\begin{align*}
\int_{1}^{\infty} x^{-\gamma}\exp\left( -\nu x^{1-\beta} \right){\rm{d}}x
=& \int_{\nu}^\infty \frac{1}{\nu(1-\beta)} \left(\frac{t}{\nu} \right)^{\frac{1-\gamma}{1-\beta}-1}e^{-t}\textrm{d}t\\
=& \frac{\nu^{-\frac{1-\gamma}{1-\beta}}}{1-\beta} \Gamma\left(\frac{1-\gamma}{1-\beta},\nu\right).
\end{align*}
This proves the lemma.
\begin{lemmax}\label{lemma:Gamma1}
For $\alpha,\ \gamma>0, \ a_{2}\geq0, \ \alpha-\gamma>1$, we have
\begin{align*}
&\int_{2}^{\infty} \frac{(x-1+a_{2})^{\gamma}}{(x-2+a_{2})^{\alpha}}{\rm{d}}x \\
\leq& \left|\sum_{t=0}^{[\gamma]}\frac{(-1)^{t}
\prod_{\varsigma=1}^{t}(\gamma-t+\varsigma)(1+a_{2})^{\gamma-t}a_{2}^{-\alpha+t+[\gamma]+1}}
{\prod_{\varsigma=1}^{t+1}(-\alpha+\varsigma)}\right|\\
&+\frac{\prod_{\varsigma=0}^{[\gamma]}(\gamma-[\gamma]+\varsigma)
a_{2}^{-\alpha+\gamma+1}}{\prod_{\varsigma=0}^{[\gamma]}
(\alpha-[\gamma]-1+\varsigma)(\alpha-\gamma-1)},
\end{align*}
where $[\gamma]$ is the largest integer not greater than $\gamma$.
\end{lemmax}
\textit{Proof}: Denote $\iota=[\gamma]+1$. Then, by the formula of multiple integration-by-parts \cite{Zorich2016}, we have
\begin{align*}
\!\!\!\!\!\!&\int_{2}^{\infty} \frac{(x-1+a_{2})^{\gamma}}{(x-2+a_{2})^{\alpha}}\textrm{d}x\\
\!\!\!\!\!\!=&\int_{2}^{\infty} (x-1+a_{2})^{\gamma}\left(\frac{(x-2+a_{2})^{-\alpha+\iota}}{
\prod_{\varsigma=1}^{\iota}(-\alpha+\varsigma)}\right)^{(\iota)}\textrm{d}x\\
\!\!\!\!\!\!=&\left.\sum_{t=0}^{\iota-1}(-1)^{t}((x-1+a_{2})^{\gamma})^{(t)}
\left(\frac{(x-2+a_{2})^{-\alpha+\iota}}{\prod_{\varsigma=1}^{\iota}(-\alpha+\varsigma)}\right)^{(\iota-1-t)}\right|_{2}^{\infty}\\
\!\!\!\!\!\!&+(-1)^{\iota}\int_{2}^{\infty} \frac{((x-1+a_{2})^{\gamma})^{(\iota)}(x-2+a_{2})^{-\alpha+\iota}}
{\prod_{\varsigma=1}^{\iota}(-\alpha+\varsigma)}\textrm{d}x\\
\!\!\!\leq& \left|\sum_{t=0}^{\iota-1}\frac{(-1)^{t}\prod_{\varsigma=1}^{t}
(\gamma-t+\varsigma)(1+a_{2})^{\gamma-t}a_{2}^{-\alpha+t+\iota}}{
\prod_{\varsigma=1}^{t+1}(-\alpha+\varsigma)}\right|\\
\!\!\!\!\!\!&+\frac{(-1)^{\iota}\gamma\cdots(\gamma-\iota+1)}
{(-\alpha+\iota)\cdots(-\alpha+1)}\int_{2}^{\infty} (x-2+a_{2})^{-\alpha+\gamma}\textrm{d}x\\
\!\!\!\!\!\!\leq& \left|\sum_{t=0}^{\iota-1}\frac{(-1)^{t}\gamma\cdots(\gamma-t+1)(1+a_{2})^{\gamma-t}
a_{2}^{-\alpha+t+\iota}}{\prod_{\varsigma=1}^{t+1}(-\alpha+\varsigma)}\right|\\
\!\!\!\!\!\!&+\frac{\gamma(\gamma-1)\cdots(\gamma-\iota+1)a_{2}^{-\alpha+\gamma+1}}{(\alpha-\iota)\cdots(\alpha-1)(\alpha-\gamma-1)}
\end{align*}
This proves the lemma. $\hfill\Box$

\begin{theorem}\label{Preepsilon}
Suppose Assumption \ref{Assumption_Graph} holds. Given a parameter $\epsilon^\star$ and let the step-size  $\alpha(k)=\frac{\alpha_k}{(k+a_{2})^{\beta}}, 0<\underline{\alpha}\leq\alpha_k\leq\overline{\alpha}<\infty$, $\beta\in(0,1]$, $b(k)=\underline{b}(k+a_{2}-1)^{\gamma}$, the algorithm is $\epsilon^\star$-differentially private as follows.\\ When $\beta\!=\!1$, $\underline{\alpha}c_{\min}\!+\!\gamma\!>\!1$, $\gamma\!\geq\!0$, we have
\begin{eqnarray}\label{givepsilon1}
\!\!\frac{\delta}{b(1)}\!\!+\!\!\frac{\delta(1+a_{2})^{\overline{\alpha} c_{\min}}a_{2}^{1-\underline{\alpha} c_{\min}-\gamma}}{\underline{b}(\underline{\alpha}c_{\min}+\gamma-1)}\leq\epsilon^\star.
\end{eqnarray}
When $\beta\!=\!1$, $\underline{\alpha}c_{\min}\!+\!\gamma\!>\!1$, $\gamma\!<\!0$, we have
\begin{eqnarray}\label{givepsilon2}
\!\!&\!\!\frac{\delta}{b(1)}\!\!+\!\!\frac{\delta(1\!+\!a_{2})^{\overline{\alpha}c_{\min}}}{\underline{b}}
\left(\left|\sum\limits_{t=0}^{[-\gamma]}\!\!
\frac{\gamma\cdots(\gamma\!+\!t\!-\!1)(1\!+\!a_{2})^{-\!\gamma\!-\!t}a_{2}^{-\underline{\alpha}c_{\min}\!+\!t\!+\![-\gamma]+1}}{(-\underline{\alpha}c_{\min}\!+\!1\!+\!t)
\cdots(-\underline{\alpha}c_{\min}\!+\!1)}\right|\right. \cr
\!\!&\!\!+\left.\!\!\frac{-\gamma\cdots(-\!\gamma\!-\![-\gamma])a_{2}^{-\underline{\alpha}c_{\min}\!-\!\gamma\!+\!1}}{(\underline{\alpha}c_{\min}\!-\![-\!\gamma]\!-\!1)
\!\cdots\!(\underline{\alpha}c_{\min}\!-\!1)(\underline{\alpha}c_{\min}\!+\!\gamma\!-\!1)}\!\right)\leq\epsilon^\star.
\end{eqnarray}
When $0<\beta<1$ and $\gamma\geq 0$, we have
\begin{align}\label{givepsilon3}
\!\!\!\!\!\frac{\delta}{b(1)}\!\!+\!\!\frac{\delta\exp\left( \frac{\overline{\alpha}c_{\min}}{1-\beta}a_{2}^{1-\beta}\!\right)\!}{\underline{b}(1\!-\!\beta)}\! \left(\!\frac{\underline{\alpha}c_{\min}}{1-\beta}\!\right)\!^{\!-\!\frac{1-\gamma}{1-\beta}} \Gamma\!(\! \frac{1\!-\!\gamma}{1\!-\!\beta} , \!\frac{\underline{\alpha}c_{\min}a_{2}^{1-\beta}}{1-\beta}\!\!)\!\!\le\!\! \epsilon^\star.
\end{align}
When $0<\beta<1$ and $\gamma<0$, we have
\begin{eqnarray}\label{givepsilon4}
\!\!&\!\!\frac{\delta}{b(1)}\!\!+\!\!\frac{\delta\exp\left( \frac{\overline{\alpha}c_{\min}}{1\!-\!\beta}a_{2}^{1\!-\!\beta}\right)}{\underline{b}(1\!-\!\beta)} \!\left(\frac{\underline{\alpha}c_{\min}}{1\!-\!\beta}\right)^{-\frac{1\!-\!\gamma}{1\!-\!\beta}} \Gamma\left(\!\frac{1\!-\!\gamma}{1\!-\!\beta}, \!\frac{\underline{\alpha}c_{\min}a_{2}^{1\!-\!\beta}}{1\!-\!\beta}\right)\cr
\!\!&+\!\!\frac{\delta\exp\left( \frac{\overline{\alpha}c_{\min}}{1\!-\!\beta}a_{2}^{1\!-\!\beta}\right)}{\underline{b}}(\frac{-\gamma}{\underline{\alpha}c_{\min}})^{\frac{-\gamma}{1\!-\!\beta}} \exp\left(\frac{\gamma}{1\!-\!\beta}\left(\frac{-\!\gamma}{\underline{\alpha}c_{\min}}\right)^{-\!\frac{1}{\beta}}\right)\!\!\le\!\! \epsilon^\star.
\end{eqnarray}
\end{theorem}
\textit{Proof}: From Theorem \ref{Sensitivity} and substituting $\alpha(k)=\frac{\alpha_k}{(k+a_{2})^{\beta}}$ into equation (\ref{TSensitivity}), we have
\begin{align*}
S(k) \le \left\{ \begin{array}{l}
\delta, ~k=1\\
 \prod_{l=0}^{k-2} (1-\frac{\alpha_l c_{\min}}{(l+a_{2})^{\beta}})\delta, ~k\ge 2.
\end{array} \right.
\end{align*}
Notice that $S(1)=\delta$. Then one can get
\begin{align*}
\epsilon = \sup_T\epsilon_T = \sum_{k=1}^\infty \frac{S(k)}{b(k)} = \frac{\delta}{b(1)}+\sum_{k=2}^\infty \frac{S(k)}{b(k)}\le \epsilon^\star.
\end{align*}
Thus, it suffices to analyze $\sum_{k=2}^\infty \frac{S(k)}{b(k)}$. The following
analysis is undertaken according to four cases.
		
\textbf{Case 1:} $\bm{\beta=1, \gamma\geq0,}$ \textbf{and} $\bm{\underline{\alpha}c_{\min}+\gamma>1}$.

From (\ref{TSensitivity}) and Lemma \ref{lemma:prod_est} it follows that
\begin{align*}
\sum_{k=2}^\infty \frac{S(k)}{b(k)}
=&\sum_{k=2}^\infty\prod_{l=0}^{k-2} (1-\frac{\alpha_kc_{\min}}{(l+a_{2})^{\beta}})\delta\nonumber\\
\leq& \sum_{k=2}^\infty \frac{\delta(1+a_{2})^{\alpha_{k}c_{\min}}}{b(k)(k-2+a_{2})^{\alpha_{k}c_{\min}}}\nonumber\\
\leq& \sum_{k=2}^\infty \frac{\delta(1+a_{2})^{\overline{\alpha} c_{\min}}}{\underline{b}(k-2+a_{2})^{\underline{\alpha}c_{\min}+\gamma}}\nonumber\\
\leq& \int_{2}^\infty \frac{\delta(1+a_{2})^{\overline{\alpha} c_{\min}}}{\underline{b}(x-2+a_{2})^{\underline{\alpha}c_{\min}+\gamma}}\text{d}x \nonumber\\
\leq& \frac{\delta(1+a_{2})^{\overline{\alpha} c_{\min}}a_{2}^{1-\underline{\alpha} c_{\min}-\gamma}}{\underline{b}(\underline{\alpha}c_{\min}+\gamma-1)}.
\end{align*}
\textbf{Case 2:} $\bm{\beta=1, \gamma<0}$, \textbf{and} $\bm{\underline{\alpha}c_{\min}+\gamma>1}$.

From (\ref{TSensitivity}), Lemmas \ref{lemma:prod_est} and \ref{lemma:Gamma1} it follows that
\begin{align*}
&\sum_{k=2}^\infty \frac{S(k)}{b(k)}\\
\leq& \sum_{k=2}^\infty \frac{\delta(1+a_{2})^{\overline{\alpha}c_{\min}}}{b(k)(k-2+a_{2})^{\underline{\alpha}c_{\min}}}\nonumber\\
\leq&\int_{2}^\infty \frac{\delta(1+a_{2})^{\overline{\alpha}c_{\min}}}{\underline{b}(x-1+a_{2})^{\gamma}(x-2+a_{2})^{\underline{ \alpha}c_{\min}}}\text{d}x\\
\leq&\!\frac{\delta\!(1\!+\!a_{2})^{\overline{\alpha}c_{\min}}}{\underline{b}}
\!\!\!\left(\left|\sum_{t=0}^{[-\gamma]}\!\frac{\!\gamma\!\cdots\!(\gamma\!+\!t\!-\!1)(1\!+\!a_{2})^{-\!\gamma\!-\!t}\!a_{2}^{-\underline{\alpha}c_{\min}\!+\!t\!+\![-\gamma]\!+\!1}}{(-\underline{\alpha}c_{\min}\!+\!1\!+\!t)
\cdots(-\underline{\alpha}c_{\min}\!+\!1)}\right|\right.\\
&+\left.\frac{-\gamma\cdots(-\gamma-[-\gamma])a_{2}^{-\underline{\alpha}c_{\min}-\gamma+1}}{(\underline{\alpha}c_{\min}-[-\gamma]-1)\cdots(\underline{\alpha}c_{\min}-1)(\underline{\alpha}c_{\min}+\gamma-1)}\right).
\end{align*}
\textbf{Case 3:} $\bm{0<\beta<1, \gamma\geq 0}$.

From Lemma \ref{lemma:prod_est} it follows that
\begin{align*}
\sum_{k=2}^\infty \frac{S(k)}{b(k)} \leq& \sum_{k=2}^\infty \frac{\delta}{\underline{b}}(k+a_{2}-1)^{-\gamma}\exp\left( \frac{\overline{\alpha}c_{\min}}{1-\beta}a_{2}^{1-\beta}\right. \nonumber\\
&-\left.\frac{\underline{\alpha}c_{\min}}{1-\beta}(k+a_{2}-1)^{1-\beta}\right)
\end{align*}
Further, when $\gamma\geq 0$, by Lemma \ref{lemma:Gamma}, we have
\begin{align}\label{epsilon41}
&\sum_{k=2}^\infty \frac{S(k)}{b(k)}\nonumber\\
\leq& \exp\left( \frac{\overline{\alpha}c_{\min}}{1-\beta}a_{2}^{1-\beta}\right)\int_1^\infty \frac{\delta}{\underline{b}}(x+a_{2}-1)^{-\gamma}\nonumber\\
&\qquad\cdot\exp\left( -\frac{\underline{\alpha}c_{\min}}{1-\beta}(x+a_{2}-1)^{1-\beta}\right) \text{d}x.\nonumber\\
	=&\frac{\delta\exp\left( \frac{\overline{\alpha}c_{\min}}{1-\beta}a_{2}^{1-\beta}\right)}{\underline{b}(1-\beta)} \left(\frac{\underline{\alpha}c_{\min}}{1-\beta}\right)^{-\frac{1-\gamma}{1-\beta}}\nonumber\\ &\qquad\cdot\Gamma\left( \frac{1-\gamma}{1-\beta}, \frac{\underline{\alpha}c_{\min}a_{2}^{1-\beta}}{1-\beta} \right).
\end{align}
\textbf{Case 4:} $\bm{0<\beta<1, \gamma<0}$.

Set
\begin{align*}
g(x)=(x+a_{2}-1)^{-\gamma}\exp\left( -\frac{\underline{\alpha}c_{\min}}{1-\beta}(x+a_{2}-1)^{1-\beta}\right).
\end{align*}
Then, we have
\begin{align*}
g^\prime(x)&=(x+a_{2}-1)^{-\gamma-1}\exp\left( -\frac{\underline{\alpha}c_{\min}}{1-\beta}(x+a_{2}-1)^{1-\beta}\right)\\
&\cdot\left(-\gamma - \underline{\alpha}c_{\min}(x+a_{2}-1)^{-\beta}\right),
\end{align*}
which implies that $ g(x)$ increases monotonically in $(0,(\frac{-\gamma}{\underline{\alpha}c_{\min}})^{\frac{-1}{\beta}}-a_{2}+1)$, and decreases monotonically in $((\frac{-\gamma}{\underline{\alpha}c_{\min}})^{\frac{-1}{\beta}}-a_{2}+1,\infty)$.

Let $k_g$ be the greatest integer less than or equal to $ \left(\frac{-\gamma}{\underline{\alpha}c_{\min}}\right)^{\frac{-1}{\beta}}-a_{2}+1$. When $k_g<2$, then we get (\ref{epsilon41}). When $k_g\geq 2$, then we get
\begin{align*}
&\sum_{k=2}^\infty(k+a_{2}-1)^{-\gamma}\exp\left(-\frac{\underline{\alpha}c_{\min}}{1-\beta}(k+a_{2}-1)^{1-\beta}\right)\\
	\leq& \sum_{k=2}^{k_g} (k+a_{2}-1)^{-\gamma}\exp\left( -\frac{\underline{\alpha}c_{\min}}{1-\beta}(k+a_{2}-1)^{1-\beta}\right)\\&
+\sum_{k_g+1}^{\infty} (k+a_{2}-1)^{-\gamma}\exp\left( -\frac{\underline{\alpha}c_{\min}}{1-\beta}(k+a_{2}-1)^{1-\beta}\right)\\
	\leq&\!\!\int_1^{k_g+1}\!\!(x+a_{2}-1)^{-\gamma}\exp\left( \!\!-\frac{\underline{\alpha}c_{\min}}{1-\beta}(x+a_{2}-1)^{1-\beta}\right) \text{d}x\\&\!\!+\int_{k_g}^\infty (x+a_{2}-1)^{-\gamma}\exp\left( \!\!-\frac{\underline{\alpha}c_{\min}}{1-\beta}(x+a_{2}-1)^{1-\beta}\right) \text{d}x\\
=& \int_1^{\infty} (x+a_{2}-1)^{-\gamma}\exp\left( -\frac{\underline{\alpha}c_{\min}}{1-\beta}(x+a_{2}-1)^{1-\beta}\right) \text{d}x\\
&\!\!+\!\!\int_{k_g}^{k_g+1}\!\!(x\!+\!a_{2}\!-\!1)^{-\gamma}\exp\left( \!\!-\frac{\underline{\alpha}c_{\min}}{1-\beta}(x+a_{2}-1)^{1-\beta}\right) \text{d}x\\
	=&\int_1^{\infty}(x+a_{2}-1)^{-\gamma}\exp\left( -\frac{\underline{\alpha}c_{\min}}{1-\beta}(x+a_{2}-1)^{1-\beta}\right) \text{d}x\\
	& + \sup_{x>1} (x+a_{2}-1)^{-\gamma}\exp\left( -\frac{\underline{\alpha}c_{\min}}{1-\beta}(x+a_{2}-1)^{1-\beta}\right)\\
	=&\frac{1}{(1-\beta)} \left(\frac{\underline{\alpha}c_{\min}}{1-\beta}\right)^{-\frac{1-\gamma}{1-\beta}} \Gamma\left( \frac{1-\gamma}{1-\beta} , \frac{\underline{\alpha}c_{\min}a_{2}^{1-\beta}}{1-\beta} \right) \\ &+\left(\frac{-\gamma}{\underline{\alpha}c_{\min}}\right)^{\frac{\gamma}{\beta}} \exp\left(\frac{\gamma}{1-\beta}(\frac{-\gamma}{\underline{\alpha}c_{\min}})^{-\frac{1}{\beta}} \right).
\end{align*}
Thus, (\ref{givepsilon4}) holds. This completes the proof.  $\hfill\Box$

\begin{remark}
Theorem \ref{Preepsilon} provides the upper bound of the differential privacy level $\epsilon$ when the step-size $\alpha(k)$ and the noise parameter $b(k)$ are designed in a certain form. From (\ref{TSensitivity}), (\ref{Tepsilon}) and (\ref{givepsilon1})-(\ref{givepsilon4}), increasing $\beta$ has the same effect as decreasing $\gamma$ on both the differential privacy level $\epsilon$ and the obtained boundary. Moreover, it is known that  $\epsilon$ increases as $\alpha(k)$ decreases, namely, $\beta$ increases (or $\alpha_k$ decreases, or $a_2$ increases). Similarly, the obtained boundary increases as $\beta$ increases (or $\alpha_k$ decreases, or $a_2$ increases).
\end{remark}
\begin{remark}
Since the step-size does not change $S(1)$, $\epsilon>1/b(1)$ is required regardless of the step-size. Given $\epsilon^\star>1/b(1), \beta\in(0,1], \gamma<1$ and $\underline{b}>0$, as long as $\alpha_k$ is large enough, it can always be $\epsilon\leq \epsilon^\star $. When $\beta<1$ or $\underline{\alpha}c_{\min}+\gamma>1$, given $\epsilon^\star>1/b(1), \beta\in(0,1], \gamma<1$ and $\alpha_k \geq \underline{\alpha}>0$, as long as $\underline{b}$ is large enough, it can always be $\epsilon\leq \epsilon^\star$.
\end{remark}

\subsection{Trade-off between convergence rate and privacy level}
Given the predefined indices $(s^\star, r^\star)$-accuracy and $\epsilon^\star$-differential privacy, and based on Theorems \ref{Preaccuracy} and \ref{Preepsilon}, we design the noise parameter $b(k)=O(k^{\gamma})$ and the step-size $\alpha(k)=\frac{a_1}{(k+a_2)^\beta}$ with $a_1,a_2>0$ and $\beta\in(0,1]$, such that (\ref{ConditionAccuracy}) and (\ref{givepsilon1})-(\ref{givepsilon4}) hold. Furthermore, we find that $\beta\in(0,1]$ and $\gamma<\beta-1/2$ can guarantee the mean-square convergence of the algorithm in Theorem \ref{BiasedConsensus}, while $\beta\in(1/2,1]$ and $\gamma<\beta-1/2$ can guarantee the almost-sure convergence of the algorithm in Theorem \ref{Theorem_Convergence3}. However, in Theorem \ref{Preepsilon}, we only require $\beta\in(0,1]$ to ensure the algorithm to be $\epsilon^\star$-differentially private. So, in order to satisfy both conditions in Theorems \ref{BiasedConsensus} and \ref{Preepsilon}, we require $\beta\in(0,1]$ and $\gamma<\beta-1/2$, and $\beta\in(1/2,1]$ and $\gamma<\beta-1/2$ for Theorems \ref{Theorem_Convergence3} and \ref{Preepsilon}.

From (\ref{them31}), (\ref{them32}), (\ref{them41}), (\ref{them42}) and (\ref{givepsilon1})-(\ref{givepsilon4}), we observe that the influence of the privacy noise on the convergence rate and the privacy level of the algorithm is different. Specifically, if the privacy noise increases ($\gamma$ increases), the convergence rate of the algorithm will slow down, but the privacy of the algorithm will be enhanced ($\epsilon$ decreases). This is intuitive because the increase of the privacy noise enhances data randomness, leading to a worse convergence rate and more robust privacy of the algorithm. These findings are consistent with the current literature's results on differentially private algorithms, i.e., there is a trade-off between the convergence and the privacy of the algorithm.
\section{Numerical example}
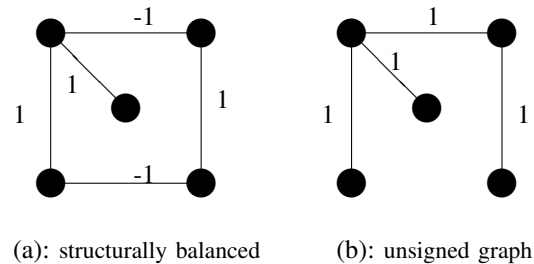
\begin{figure}[!ht]
\begin{center}
\setlength{\unitlength}{1mm}
\begin{picture}(80, 40)
\put(5,10){\circle*{4}}
\put(25,10){\circle*{4}}
\put(15,20){\circle*{4}}
\put(5,30){\circle*{4}}
\put(25,30){\circle*{4}}

\put(0,18){1}
\put(16,10){-1}
\put(7,22){1}
\put(16,31){-1}
\put(27,20){1}

\put(5,10){\line(0,1){20}}
\put(5,10){\line(1,0){20}}
\put(5,30){\line(1,0){20}}
\put(25,10){\line(0,1){20}}
\put(15,20){\line(-1,1){10}}
\put(0,0){(a): {\small structurally balanced}}
\put(45,10){\circle*{4}}
\put(65,10){\circle*{4}}
\put(55,20){\circle*{4}}
\put(45,30){\circle*{4}}
\put(65,30){\circle*{4}}
\put(41,18){1}
\put(67,18){1}
\put(50,25){1}
\put(55,31){1}
\put(45,30){\line(0,-1){20}}
\put(45,30){\line(1,0){20}}
\put(45,30){\line(1,-1){10}}
\put(65,10){\line(0,1){20}}
\put(43,0){(b): {\small unsigned graph}}
\end{picture}
\end{center}
\caption{Communication topology} \label{Fig_Graph}
\end{figure}

This section considers discrete-time multi-agent systems of five agents coupled by the communication graph illustrated in Fig. \ref{Fig_Graph}. In this example, we aim to achieve bipartite consensus with the $(s^\star,r^\star)$-accuracy and $\epsilon^\star$-differential privacy, where $s^\star=0.59$, $r^\star=9$, and $\epsilon^\star$=2.5.
\begin{figure*}[htbp]
	\centering
	\subfigure[Algorithm: $\alpha(k)=0.3/(k+1)$, $b(k)=k^{0.1}$]{
		\begin{minipage}[b]{0.45\textwidth}
			\includegraphics[width=1\textwidth,height=0.62\textwidth]{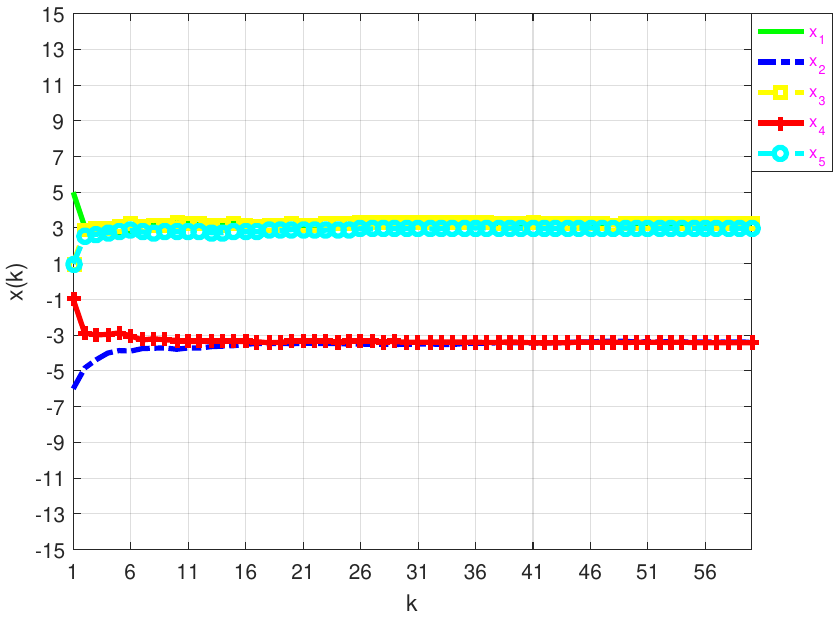}
			\includegraphics[width=1\textwidth,height=0.65\textwidth]{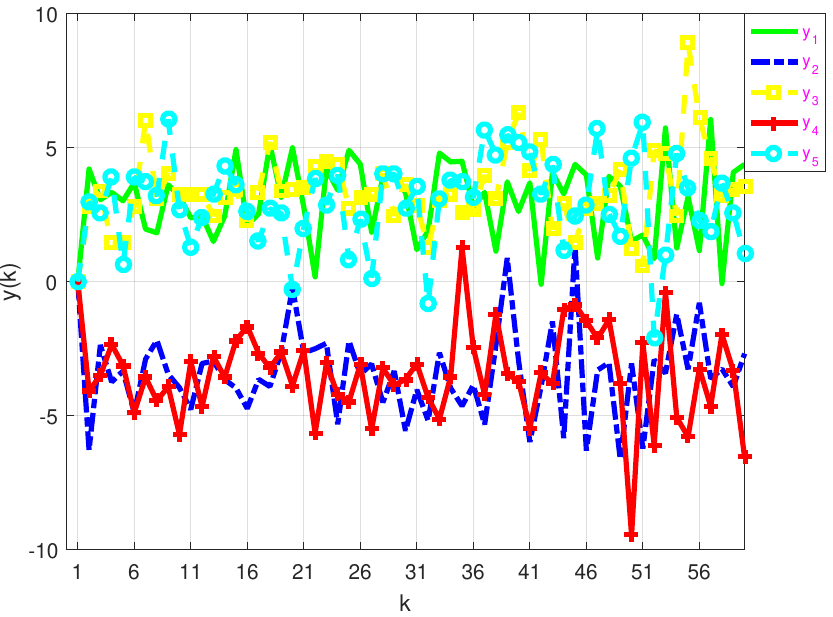}
		\end{minipage}}
	\subfigure[\cite{Zuo2022}: $c_i=1$, $q_i=0.9$, $p_{i}=0.8$]{
		\begin{minipage}[b]{0.45\textwidth}
			\includegraphics[width=1\textwidth,height=0.65\textwidth]{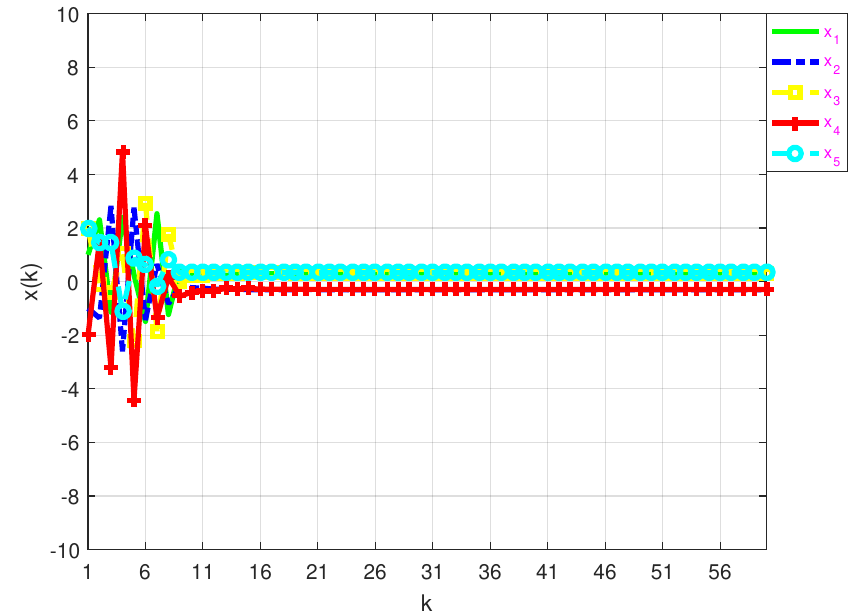}
			\includegraphics[width=0.95\textwidth,height=0.63\textwidth]{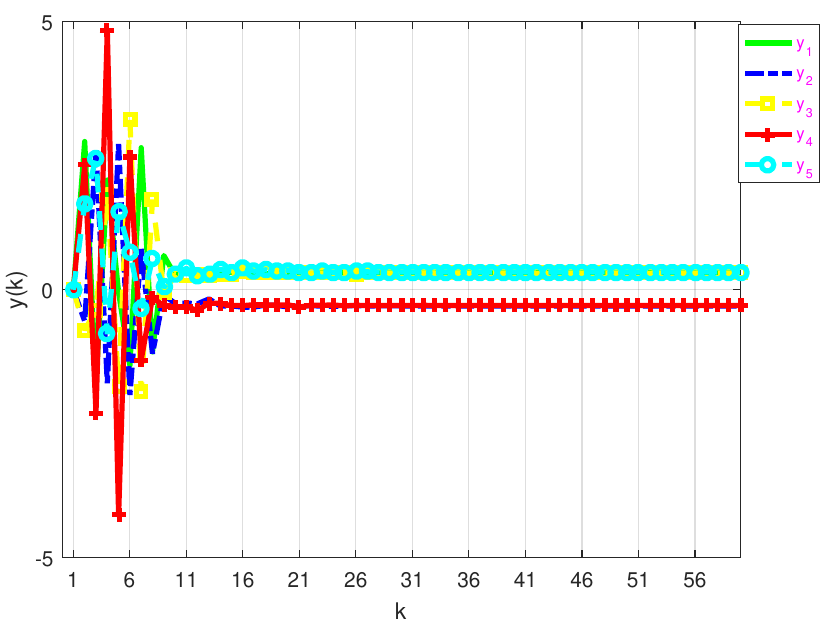}
		\end{minipage}}
\caption{Comparison of the algorithm with existing privacy-preserving approach in \cite{Zuo2022}}
\label{Fig_XiTra}
\end{figure*}
\begin{figure*}[htbp]
\subfigure[Algorithm: $\alpha(k)=1/(k+1)^{0.9}$, $b(k)=k^{0.1}$]{
		\begin{minipage}[b]{0.45\textwidth}
		\includegraphics[width=1\textwidth,height=0.63\textwidth]{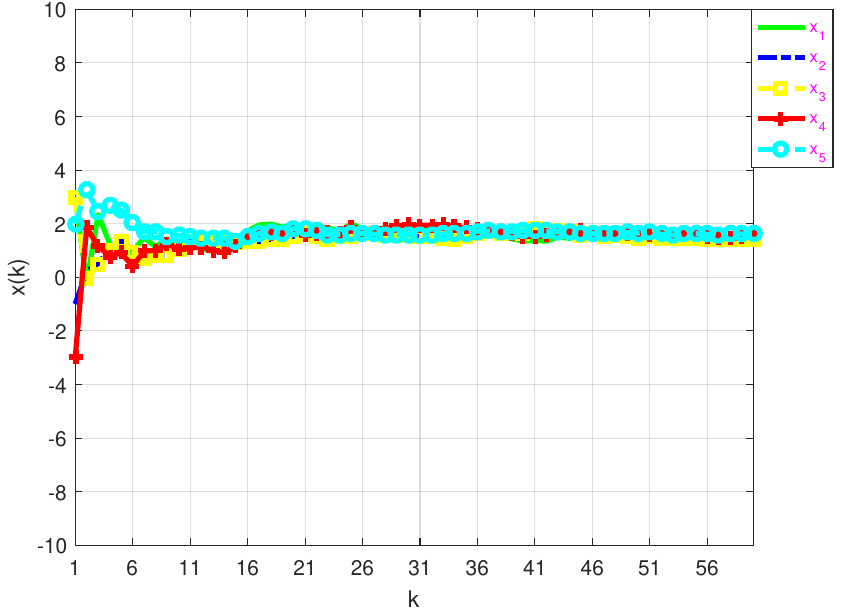}
		\includegraphics[width=1\textwidth,height=0.63\textwidth]{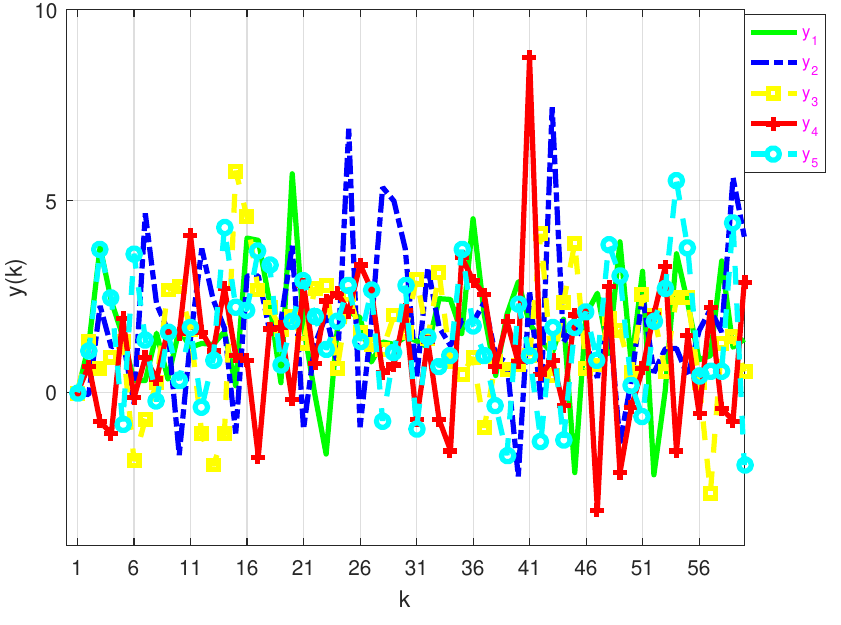}
		\end{minipage}}
\subfigure[\cite{Nozari2017Differentially}: $c_i=1$, $q_i=0.9$, $p_{i}=0.8$]{
		\begin{minipage}[b]{0.45\textwidth}
		\includegraphics[width=1\textwidth,height=0.65\textwidth]{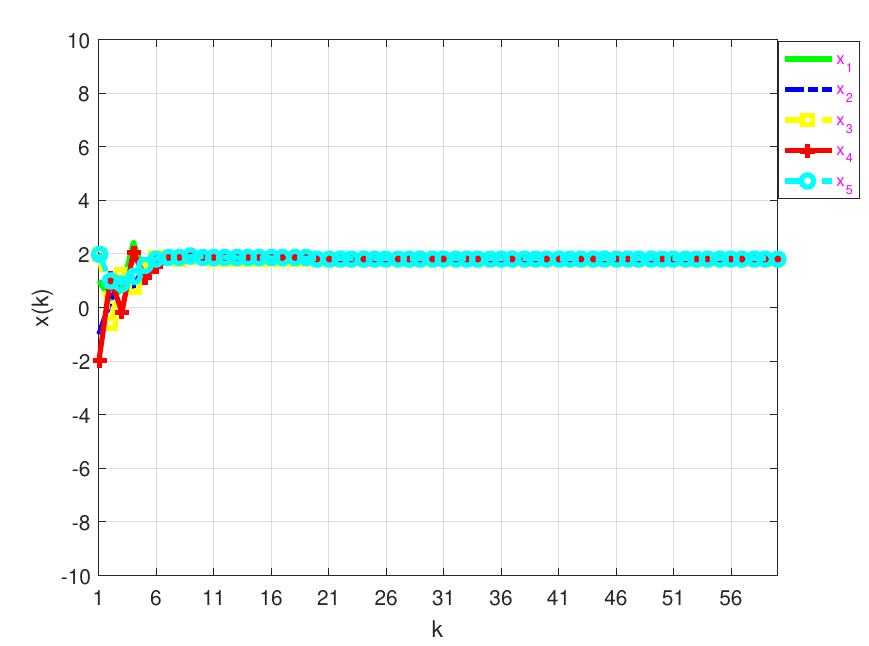}
	    \includegraphics[width=1\textwidth,height=0.65\textwidth]{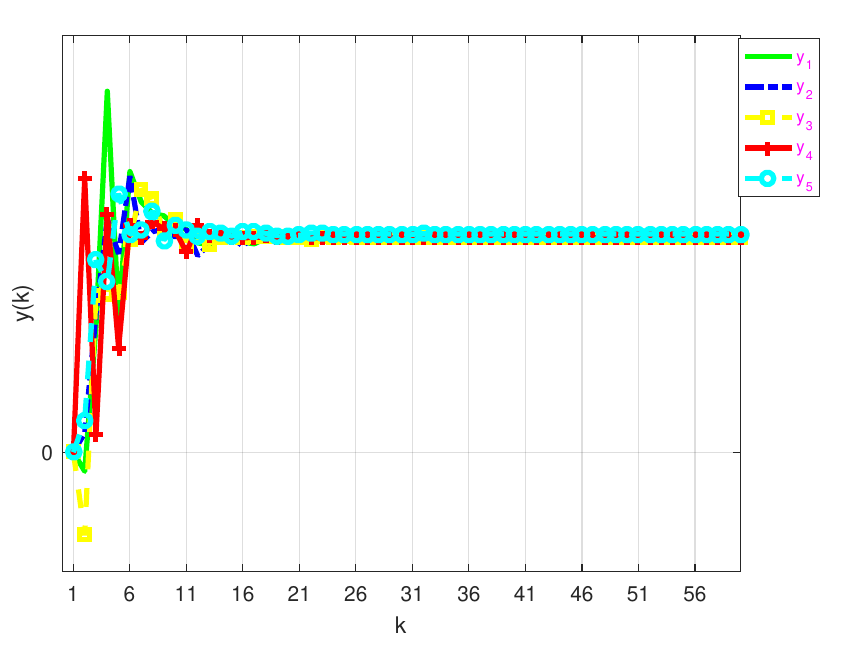}
		\end{minipage}}
\caption{Comparison of the algorithm with existing privacy-preserving approach in \cite{Nozari2017Differentially}}
\label{Fig_XiTra1}
\end{figure*}

First, for the communication topology (a) in Fig. \ref{Fig_Graph}, we employ the proposed controller (\ref{controller}) using a increasing variance of the privacy noises (\ref{NoiseAdd}) with $b(k)=k^{0.1}$, and compare to decaying variance of the privacy noises with $c_i=1$, $q_i=0.9$ in \cite{Zuo2022}. Considering the increasing variance of the privacy noises, we set the step-size as $\alpha(k) =0.5/(k+1)$, while for the decaying variance of the privacy noises, we adopt the method in \cite{Zuo2022} and set the step-size as $\alpha(k)=0.8^{k}$. The corresponding results are depicted in Fig. \ref{Fig_XiTra} (a) and (b), respectively, where the trajectories of $x(k)$ and $y(k)$  are displayed. The former figure in Fig. \ref{Fig_XiTra} (a) and (b) reveals that both algorithms of the paper and \cite{Zuo2022} converge. The latter figure in Fig. \ref{Fig_XiTra} (a) and (b) reveals that $y(k)$ utilizing the algorithm of the paper is random, while the corresponding $y(k)$ utilizing the algorithm of \cite{Zuo2022} converges.

Second, for distributed consensus with unsigned graph ($S=I$), i.e., the communication topology (b) in Fig. \ref{Fig_Graph}, the comparison between the algorithm  and \cite{Nozari2017Differentially} is illustrated in Fig. \ref{Fig_XiTra1} (a) and (b), respectively. The former figure in Fig. \ref{Fig_XiTra1} (a) and (b) also reveals that both algorithms of the paper and \cite{Nozari2017Differentially} converge. The latter figure in Fig. \ref{Fig_XiTra1} (a) and (b) also reveals that $y(k)$ utilizing the algorithm of the paper is random, while the corresponding $y(k)$ utilizing the algorithm of \cite{Nozari2017Differentially} converges.

Based on the above analysis, Fig. \ref{Fig_XiTra} and \ref{Fig_XiTra1} highlight that our algorithm has better privacy protection with guaranteed convergence, which is consistent with theoretical analysis.

\section{Conclusion}

This paper develops a differentially private bipartite consensus algorithm over signed networks. We relax the selection of privacy noises in the existing mechanisms, such that the variances of the privacy noises are time-varying and allowed to be increased with time. By using the stochastic approximation method, the proposed algorithm achieves asymptotically unbiased mean-square and almost-sure bipartite consensus, and at the same time, protects the initial value of each agent. Furthermore, we develop a method to design the time-varying step-size and the noise parameter to guarantee the desired consensus accuracy and predefined differential privacy level. We also give the mean-square and almost-sure convergence rate of the algorithm. Finally, we reveal the trade-off between the convergence rate of the algorithm and privacy level $\epsilon$ with different forms of the privacy noises. It is worth mentioning that many interesting topics deserve further investigation, including differentially private consensus-based optimization over signed networks and realizing privacy security for multi-agent systems under active adversaries.

\end{document}